\documentclass[journal]{IEEEtran}
\usepackage{amsmath,amsfonts}
\usepackage{algorithmic}
\usepackage{algorithm}
\usepackage{array}
\usepackage[caption=false,font=normalsize,labelfont=sf,textfont=sf]{subfig}
\usepackage{textcomp}
\usepackage{stfloats}
\usepackage{url}
\usepackage{verbatim}
\usepackage{graphicx}
\usepackage{cite}
\usepackage{xcolor}         
\usepackage{tcolorbox}
\usepackage{booktabs}
\usepackage{hyperref}
\usepackage{pifont}
\usepackage{balance}

\begin{document}

\title{MAESTRO: a Multimodal Auditory-attention Egocentric Speech-TRacking Open corpus}

\author{K~M~Naimul~Hassan\textsuperscript{*},~\IEEEmembership{Student Member,~IEEE},
        Ali~Alavi\textsuperscript{*},~\IEEEmembership{Student Member,~IEEE},
        and~Donald~S.~Williamson,~\IEEEmembership{Senior Member,~IEEE}\\
\thanks{\textsuperscript{}}
\thanks{This work has been submitted to the IEEE for possible publication. Copyright may be transferred without notice, after which this version may no longer be accessible.}
\thanks{Corresponding authors: K M Naimul Hassan and Ali Alavi. \\
K M Naimul Hassan, Ali Alavi, and Donald S. Williamson are with the Department of Computer Science and Engineering, The Ohio State University, Columbus, OH 43210 USA (e-mail: hassan.491@osu.edu, alavibajestan.1@osu.edu, williamson.413@osu.edu). Donald S. Williamson is also with the Center for Cognitive and Brain Sciences, The Ohio State University.}
\thanks{\textsuperscript{*}Equal contribution.}
}

\IEEEpubidadjcol

\maketitle

\begin{abstract}

Humans rely on gaze, head movements, and visual cues to attend to speakers in noisy environments, yet auditory attention decoding (AAD) has been studied primarily using electroencephalography (EEG). We introduce the Multimodal Auditory-attention Ego-centric Speech-TRacking Open (MAESTRO) corpus, the first AAD dataset to simultaneously record EEG, eye gaze, pupillometry, egocentric video, and head inertial measurement unit (IMU) data. MAESTRO includes four competing speakers and background noise across multiple signal-to-noise ratio (SNR) conditions, enabling attention decoding under realistic listening scenarios. Through a four-speaker attention decoding benchmark, we show that combining behavioral and physiological signals improves decoding performance over EEG-only approaches, enabling future advances in multimodal auditory attention decoding. These findings open the door to new applications, analyses, and methodological advances in multimodal AAD. The complete dataset is publicly available at: \href{https://huggingface.co/datasets/aspire-osu/maestro-eeg-dataset}{https://huggingface.co/datasets/aspire-osu/maestro-eeg-dataset}. The official code repository is available at: \href{https://github.com/ASPIRE-OSU/MAESTRO}{https://github.com/ASPIRE-OSU/MAESTRO}. 
\end{abstract}


\section{Introduction}
\label{sec:intro}

Auditory attention decoding~(AAD)~seeks to identify the attended speaker from neural activity, with the long-term goal of enabling hearing prostheses and intelligent hearing aids that selectively amplify desired speech. Thus, AAD addresses the cocktail party problem, which refers to the human ability to selectively attend to a single speaker in noisy, multi-talker environments \cite{cherry1953}. Most AAD research has relied on electroencephalography (EEG), a non-invasive technique that measures neural responses to auditory stimuli with high temporal resolution \cite{geirnaert2021, biesmans2016, fuglsang2017, osullivan2015, yang2026deep, ciccarelli2019}. Several datasets have driven progress in AAD. KUL \cite{biesmans2016} and DTU \cite{fuglsang2017} established the two-speaker envelope reconstruction paradigm that remains the field's primary benchmark. NJU \cite{nju2023} investigated the effects of speaker location across multiple spatial configurations, while ASA \cite{lin2024} and AASD \cite{wang2026open} expanded the problem to 10  speaker locations. ESAA \cite{esaa2022} broadened coverage to Mandarin, and MM-AAD \cite{mmaad2025} and Cocktail Party \cite{jaha2020visual} incorporated audiovisual stimuli, demonstrating that visual cues can improve performance.

These datasets have driven significant advances in AAD, but they also have key limitations. Most use dichotic two-speaker speech presentations ($\pm 90^{\circ}$), contain little or no background noise or reverberation, do not vary the signal-to-noise ratio (SNR), and often rely on non-English stimuli (Table~\ref{tab:comparison}). Consequently, models trained on these datasets may not generalize well to the noisy, reverberant, and dynamic conditions encountered by real-world hearing prostheses, particularly in English-speaking environments.

\IEEEpubidadjcol 

In real-world listening, auditory attention is supported by coordinated auditory, visual, and motor behaviors. Listeners naturally direct their gaze toward the attended speaker, adjust their head position to improve spatial hearing, and use visual cues such as lip movements and speaker location to disambiguate competing voices \cite{best2023effect, gehmacher2024eye, lertpoompunya2024head, wallach1940role, sumby1954visual, ahmed2023integration, xu2026utilizing}. Consistent with this, auditory attention has been linked to visual cortical activity \cite{golumbic2013}, gaze behavior \cite{hendrikse2019}, head orientation \cite{lertpoompunya2024head}, and pupil dilation, which reflects listening effort during speech perception in noise \cite{dimitrijevic2019neural, seifi2020exploratory}. Together, these findings suggest that auditory attention is inherently multimodal and that EEG alone may not fully capture the underlying attentional state.

Nevertheless, most existing datasets record EEG in isolation, often ignoring or suppressing natural physiological signals and behaviors by instructing participants to fixate on a central crosshair and minimize blinking \cite{esaa2022, lin2024}. Rotaru \textit{et al.}~\cite{rotaru2024} showed that spatial AAD methods that decode left versus right attention may be confounded by gaze shifts toward the attended speaker, suggesting that some EEG-based decoders may partially exploit gaze-related neural signals rather than purely auditory responses. Although MM-AAD \cite{mmaad2025} and Cocktail Party \cite{jaha2020visual} demonstrated benefits from audiovisual information, they present controlled visual stimuli rather than recording participants' natural behavior. Consequently, these datasets do not capture spontaneous gaze shifts, head movements, or egocentric visual experiences. Recording such signals alongside EEG would enable richer multimodal decoders and provide a stronger basis for evaluating the contributions of neural and behavioral cues to auditory attention decoding.

To address these limitations, we introduce the Multimodal Auditory-attention Ego-centric Speech-TRacking Open (MAESTRO) corpus, a 16-subject, 100-trial dataset that synchronously records 32-channel EEG, binocular gaze, pupillometry, egocentric video, and head inertial measurement unit (IMU) data. MAESTRO features four competing English speakers from LibriSpeech \cite{panayotov2015librispeech}, background noise from CHiME-Home \cite{foster2015chime}, a naturally reverberant recording environment, and attended-speaker SNRs ranging from 0 to 18 dB. Unlike prior datasets, participants were free to move their eyes and head naturally, enabling the capture of realistic behavioral cues associated with auditory attention. MAESTRO also records the naturally mixed acoustic scene through the egocentric camera's audio channel, reflecting the speech and noise mixture experienced by the listener. To the best of our knowledge, MAESTRO is the first AAD dataset to capture EEG, gaze, pupillometry, egocentric video, head motion, and environmental audio simultaneously, providing a platform for studying how neural and behavioral signals jointly encode auditory attention in realistic cocktail-party environments.

\begin{table*}[t]
\centering
\caption{Comparison of notable AAD datasets. \textnormal{\textit{Speech Sources}} denotes the number of competing speakers, while \textnormal{\textit{Noise Sources}} denotes dedicated background-noise sources separate from the speech streams. \textnormal{\textit{Adtl.~Modalities}} refers to data streams recorded alongside EEG, and \textnormal{\textit{Duration}} reports the total recording time (hours). \textnormal{\textit{Azimuth}} indicates loudspeaker positions relative to the listener's frontal axis. \textnormal{\textit{Room Reverberation}} describes the recording environment: \textnormal{\textit{Low}} indicates minimal reverberation, \textnormal{\textit{Simulated}} denotes artificially added reverberation, and \textnormal{\textit{Natural}} denotes real-world acoustic reflections.}
\label{tab:comparison}
\resizebox{\linewidth}{!}{%
\begin{tabular}{rcccccccccc}
\toprule
\textbf{Dataset} & \textbf{Subjects} & \textbf{EEG Ch.} &
\textbf{Speech Sources} & \textbf{Noise Sources} &
\textbf{Language} & \textbf{SNRs} & \textbf{Adtl.~modalities} & \textbf{Duration (hours)} & \textbf{Azimuth} & \textbf{Room reverberation}\\
\midrule
KUL~\cite{biesmans2016}  & 16 & 64 & 2  & \texttimes  & Dutch    & 0~dB & \texttimes & 19.2 & $\mathbf{\pm 90^\circ}$ & \texttimes \\
DTU~\cite{fuglsang2017}  & 18 & 64 & 2  & \texttimes  & Danish   & 0~dB & \texttimes & 19 & $\mathbf{\pm 60^\circ}$ & Simulated \\
NJU~\cite{nju2023} & 21 & 32 & 2  & \texttimes  & Mandarin  & 0~dB & \texttimes & 22.4 & $\mathbf{\pm\{15^\circ, 30^\circ, 45^\circ\}}$ & Low \\
 & & & & & & & & & $\mathbf{\pm\{60^\circ, 90^\circ, 120^\circ, 135^\circ\}}$ & \\
\vspace{1pt}
ESAA~\cite{esaa2022}     & 20 & 64 & 2  & \texttimes  & Mandarin & 0~dB & \texttimes & 12.7 & $\mathbf{\pm 90^\circ}$ & \texttimes \\
ASA~\cite{lin2024}       & 20 & 64 & 2  & \texttimes  & Mandarin  & 0~dB & \texttimes & 8 & $\mathbf{\pm\{5^\circ, 30^\circ\}}$ & \texttimes \\
 & & & & & & & & & $\mathbf{\pm\{45^\circ, 60^\circ, 90^\circ\}}$ &  \\
\vspace{1pt}
MM-AAD~\cite{mmaad2025}  & 50 & 32 & 2  & \texttimes  & Mandarin  & 0~dB & Video Stimuli & 45.83 & $\mathbf{\pm 90^\circ}$ & \texttimes \\
Cocktail Party~\cite{jaha2020visual} & 19 & 64 & 2 & \texttimes & English & 0~dB & Video Stimuli & 2.3 & $\mathbf{0^\circ}$ & \texttimes \\
AASD~\cite{wang2026open} & 18 & 64 & 2 & \texttimes & Mandarin & 0~dB & \texttimes & 21.0 & $\mathbf{\pm 90^\circ}$ & \texttimes \\
\textbf{MAESTRO (ours)}  & \textbf{16} & \textbf{32} & \textbf{4}
& \textbf{2} & \textbf{English}
& \textbf{0--18~dB} & \textbf{Gaze, Scene Video, IMU} & 13.3 & $\mathbf{\pm 22.5^\circ, \pm 67.5^\circ, \pm 135^\circ}$ & Natural \\
\bottomrule
\end{tabular}}
\end{table*}

Furthermore, most prior work evaluate AAD by determining which of two speakers a listener is attending to using envelope-based correlation metrics \cite{biesmans2016, fuglsang2017, accou2023decoding}. This provides limited insight into the contribution of non-EEG modalities and offers no standardized framework for comparing neural and behavioral decoding strategies. To address this gap, we define a benchmark task that identifies the attended speaker among four simultaneously presented speakers. The four-class formulation is both harder and more informative than the binary one. It lowers the prior probability of each class, and it requires the decoder to distinguish individual speakers rather than sides of space, a judgment that can be made from coarse lateralization cues alone.
\begin{figure}[b]
    \centering
    \includegraphics[width=\linewidth]{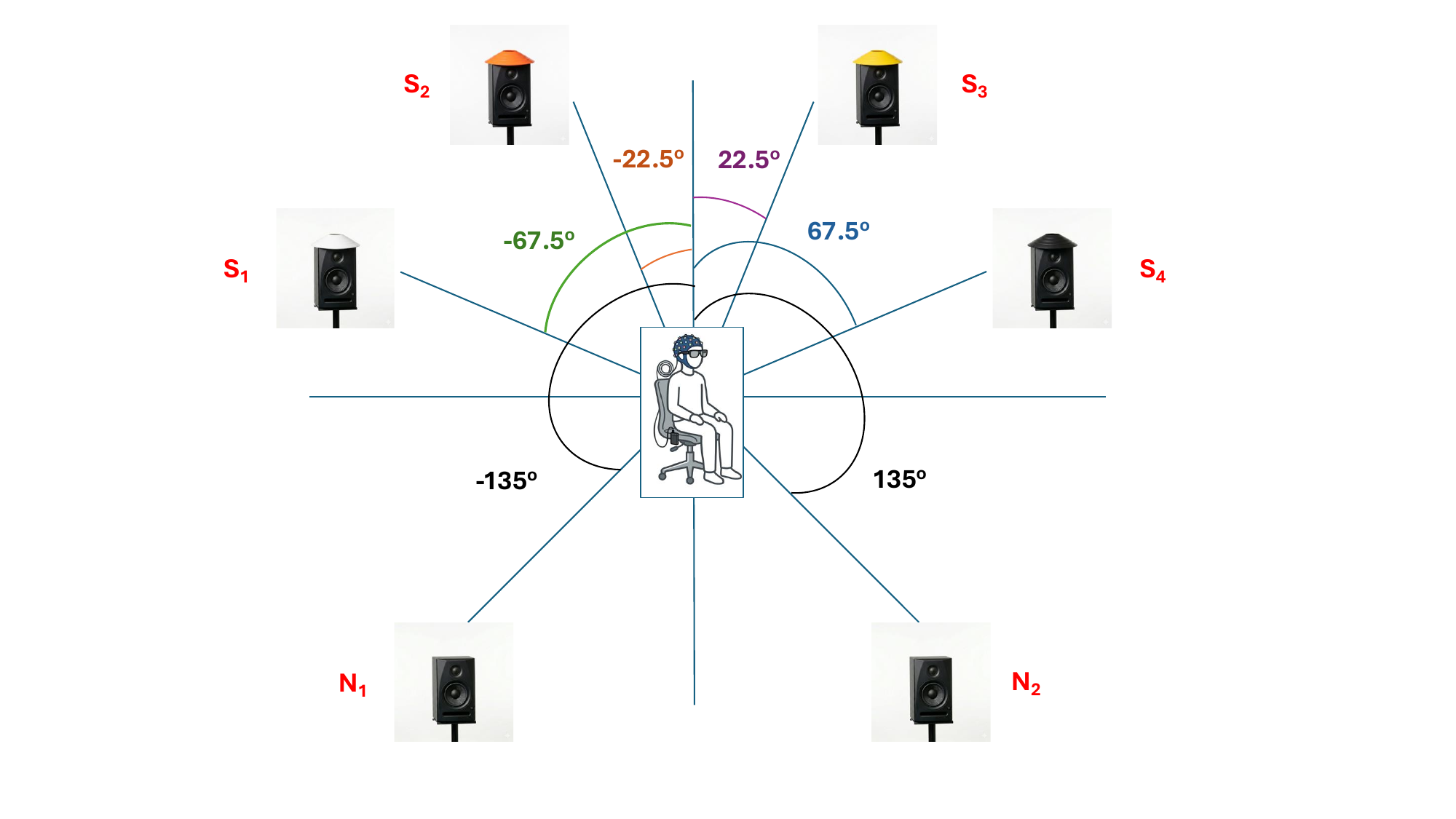}
    \caption{The spatial arrangement of loudspeakers around the participant. Four loudspeakers ($S_1$-$S_4$) were positioned relative to the participant's frontal axis, while two noise speakers ($N_1$ and $N_2$) were placed behind the participant. The participant wore an EEG headset and eye-tracking glasses.}
    \label{fig:setup}
\end{figure}
We provide a baseline evaluated across all 15 input configurations supported by MAESTRO, including individual modalities (EEG, gaze, IMU, and video), all pairwise and three-way combinations, and full multimodal fusion. We adopt a multi-encoder dilated convolutional network inspired by Accou \textit{et al.}~\cite{accou2021modeling, accou2021predicting}, a well-established architecture within the AAD literature. Rather than pursuing state-of-the-art performance, the baseline is intended to demonstrate that each modality contains decodable information and to establish reference points for future research. Results show that incorporating behavioral signals alongside EEG consistently improves the performance of the baseline.

The remainder of the paper is organized as follows. Section~\ref{sec:dataset} describes the dataset and multimodal data collection process. Section~\ref{sec:behavioral} presents a behavioral analysis of comprehension accuracy, eye gaze, and head movement patterns. Section~\ref{benchmark} introduces the evaluation protocol and the baseline, while Section~\ref{results} reports the experimental results. Section~\ref{discussion} discusses the findings, implications, limitations, and future directions, and Section~\ref{conclusion} concludes the paper.

\section{The MAESTRO Dataset} \label{sec:dataset}

\subsection{Study Configuration}

Participants were seated in a quiet, controlled room and surrounded by six loudspeakers (Fig.~\ref{fig:setup}). Four loudspeakers ($S_1$-$S_4$) were positioned at $\pm 67.5^{\circ}$ and $\pm 22.5^{\circ}$ relative to the participant's frontal axis to enable attention decoding at multiple spatial scales (left/right, near/far, and speaker identity). Two additional loudspeakers ($N_1$ and $N_2$) were positioned behind the participant at $\pm 135^{\circ}$ to provide background noise and increase acoustic complexity. All loudspeakers were located 4 ft from the participant at a height of 45.1 inches.

Data collection was managed through a custom Python graphical user interface built with \textit{PySide6}. The study comprised 105 trials: five training trials to familiarize participants with the procedure, stimuli, and interface, followed by 100 experimental trials divided into two 25-minute sessions of 50 trials each, separated by a mandatory 10-minute break (with additional breaks provided as needed to minimize fatigue). During each trial, speech and noise stimuli were presented through the loudspeakers, with each source identified by a color-coded visual cue. The interface indicated the attended speaker ($S_1$-$S_4$) and trial number, and the attended speaker was balanced across positions to prevent positional bias. Throughout the experiment, participants wore EEG, eye-tracking, and IMU sensors to simultaneously capture neural and behavioral correlates of auditory attention (see Section~\ref{sec:data_stream} for details).

\subsection{Participant Recruitment and Screening}

Sixteen participants took part in the study, with a mean age of 23.4 years (range: 18-30 years). The cohort included 9 women and 7 men, of whom 14 were right-handed and 2 were left-handed. Participants were required to be at least 18 years old and native English speakers. Participants were recruited through electronic and printed advertisements and received monetary compensation. The study was approved by the Institutional Review Board, and all participants provided written informed consent.

Prior to enrollment, participants completed four screening procedures. Cognitive function was assessed using the Mini-Mental State Examination \cite{folstein1975mini}, with a minimum score of 25 out of 30 required. Hearing was evaluated using ReSound's online hearing test \cite{resound_hearing_test}, requiring pure-tone averages across 0.5, 1, and 2 kHz to fall between 15 and 55 dB HL. Participants were also screened for allergies to the conductive gel used with the ANTNeuro EEG headset. Those who wore prescription glasses used the Tobii Pro Glasses 3 with interchangeable prescription lens inserts ranging from~$-8.0$ to~$+3.0$ diopters in 0.5-diopter increments. Screening data were not retained after eligibility determination. Participants  completed a demographic questionnaire collecting age, gender, race/ethnicity, handedness, and ear preference, after which they were assigned a unique eight-character identifier for anonymization.

\subsection{Audio Stimuli and Comprehension Questions}

Speech stimuli came from the LibriSpeech corpus \cite{panayotov2015librispeech}, a large-scale collection of English audiobook recordings chosen for its speaker diversity, high recording quality, and open license. Background noise came from the CHiME-Home dataset \cite{foster2015chime}, which contains realistic domestic soundscapes representative of everyday listening environments. Each trial included four unique 30-second speech signals presented through the front loudspeakers and two unique 30-second noise signals presented through the rear loudspeakers, yielding 630 unique audio files from 420 unique speakers across 105 trials. No audio file or speaker was repeated, preventing participants from relying on familiarity with previously heard content. We release all loudspeaker streams and the naturally mixed audio recording to facilitate downstream speech processing applications beyond the benchmark presented here.

The attended speaker's SNR varied from 0 to 18 dB and was approximately normally distributed across trials (mean~$\approx$~12 dB with a standard deviation of 3.16 dB). This distribution was chosen to reflect real-world conversational environments, where moderate SNRs are more common than extreme conditions \cite{pearsons1977speech, smeds2015estimation}. SNR was controlled on a trial-by-trial basis by scaling audio signals before playback to achieve the target ratio between the attended speaker and the combined competing speech and noise sources. Target SNR values were verified using a sound pressure meter.

Following each trial, participants answered a multiple-choice comprehension question about the attended speaker to verify attentional compliance. Questions were generated automatically using the GPT-4o API \cite{hurst2024gpt} and conditioned on the transcript of the attended speech stimulus using the following prompt:
\begin{tcolorbox}[colback=gray!5, colframe=black, fontupper=\footnotesize]
\textbf{Prompt:} Generate a multiple-choice question based on the following transcription. The question needs to be simple and short. The transcription will be read aloud, and a person will listen to it. He needs to put his attention to answering the question. So, the question is a test of his auditory attention. Provide a question, the correct answer, and three incorrect options. It is okay if the incorrect choices are not related to the transcription.
\end{tcolorbox}
\noindent
This automated approach ensured consistent question style and difficulty across trials while substantially reducing manual effort. All generated questions and answer choices were subsequently reviewed for consistency and relevance before use. An example of a stimulus transcript and its corresponding generated question is shown below.
\begin{tcolorbox}[colback=gray!5, colframe=black, fontupper=\footnotesize]
\textbf{Transcript (30s excerpt):} \textit{``The caterpillar and Alice looked at each other for some time in silence. At last the caterpillar took the hookah out of its mouth, and addressed her in a languid, sleepy voice. `Who are you,' said the caterpillar. This was not an encouraging opening for a conversation. Alice replied, rather shyly. I hardly know, sir, just\ldots''} \\[6pt]
\textbf{Question:} What does the caterpillar ask Alice? \\[4pt]
\textbf{A.} Why she is silent \\
\textbf{B.} Who she is \hspace{1em} \textit{(correct)} \\
\textbf{C.} What she wants \\
\textbf{D.} Where she is going \\
\textbf{E.} I could not pay attention
\end{tcolorbox}
\noindent
The question included five randomized response options: one correct answer, three distractors, and a dedicated~\textit{``I could not pay attention''} option. This last option was included to identify trials with complete attentional failure. Per-trial correctness labels and metadata are released alongside the dataset, enabling downstream analyses to filter trials based on attentional compliance and other desired factors.

\subsection{Multimodal Sensory Data Streams} \label{sec:data_stream}

During the experiment, EEG, gaze, egocentric video, and IMU data were recorded simultaneously across all trials. Unless otherwise noted, each modality was low-pass filtered, resampled to a common rate of 64 Hz, and z-score normalized on a per-channel, per-trial basis.
Temporal alignment across modalities was achieved through software-based synchronization. The Unix timestamp of the first recorded sample was captured independently for each modality: EEG via \textit{Lab Streaming Layer (LSL)}, gaze and IMU via the Tobii Pro Glasses~3 \textit{RTSP} stream, and audio via playback start timestamps. All recordings were acquired on a single host machine, so they shared a common system clock. During offline processing, signals were aligned to the onset of audio playback by trimming each modality according to the offset between its first-sample timestamp and the earliest audio playback timestamp.

\subsubsection{EEG}

EEG was recorded continuously at 500 Hz using 32 electrodes arranged in a standard 10-20 montage and acquired with the ANTNeuro eego mylab amplifier using a linked-mastoid reference. Acquisition quality was verified using the hardware sample counter, which showed uniform 2.0 ms sample intervals with no detected gaps or dropped samples. Raw EEG signals were notch filtered at 60 Hz and bandpass filtered from 1-40 Hz. Channels were flagged as bad if they were flat, saturated, or exhibited outlier variance relative to other channels. Data were re-referenced to the average of the available mastoid electrodes (M1, M2) when at least one mastoid channel was valid; otherwise, an average reference across all channels was used. Bad channels were then repaired using spherical spline interpolation, yielding 32 valid channels per trial, before downsampling to 64 Hz.

\begin{figure}[t]
    \centering
    \includegraphics[width=\linewidth]{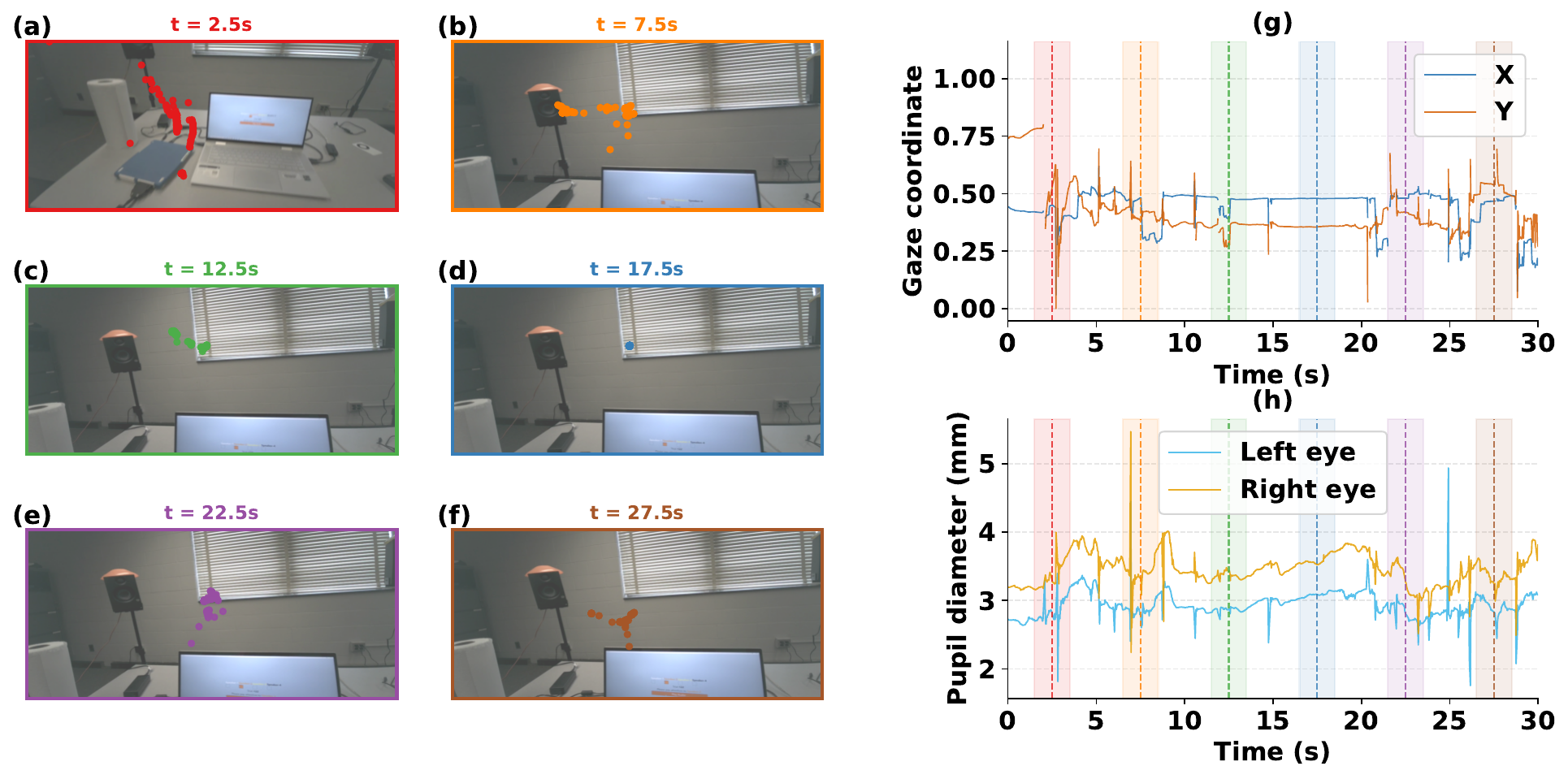}
    \caption{Sample gaze recording from Subject S02, Trial \texttt{eval\_022}. Panels (a)-(f) show six evenly spaced video frames with gaze points from a $\pm$1-second window overlaid as colored dots, whose border colors match the shaded regions in (g) and (h). Panel (g) shows horizontal (X) and vertical (Y) gaze coordinates over the trial, illustrating smooth pursuit, saccadic transitions, and occasional tracking loss; (h) shows left and right pupil diameters, bilaterally symmetric with transient dilation events.}
    \label{fig:gaze_sample}
\end{figure}

\begin{figure*}[b]
    \centering
    \subfloat[]{
        \includegraphics[width=0.48\textwidth]{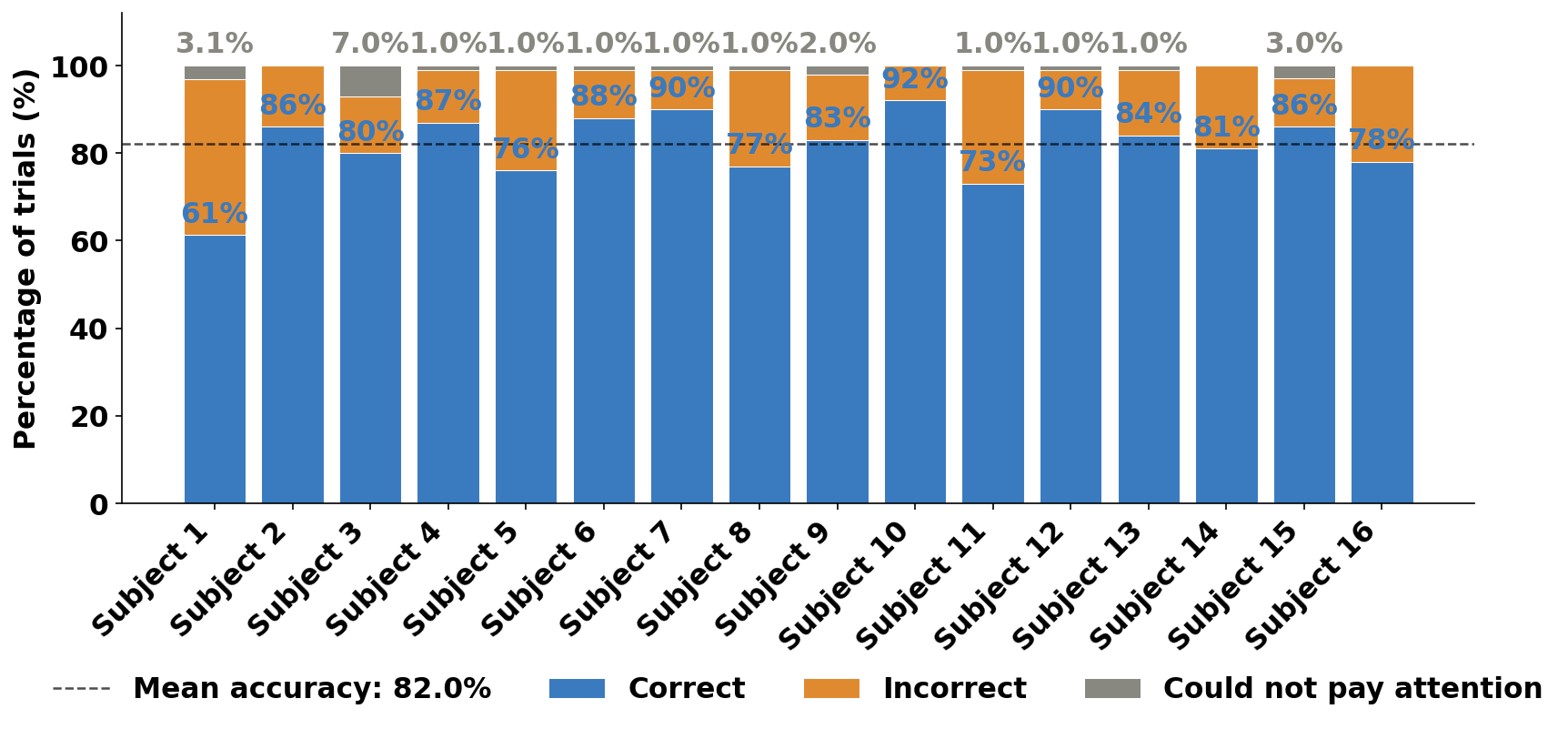}
        }
    \hfill
    \subfloat[]{
        \includegraphics[width=0.48\textwidth]{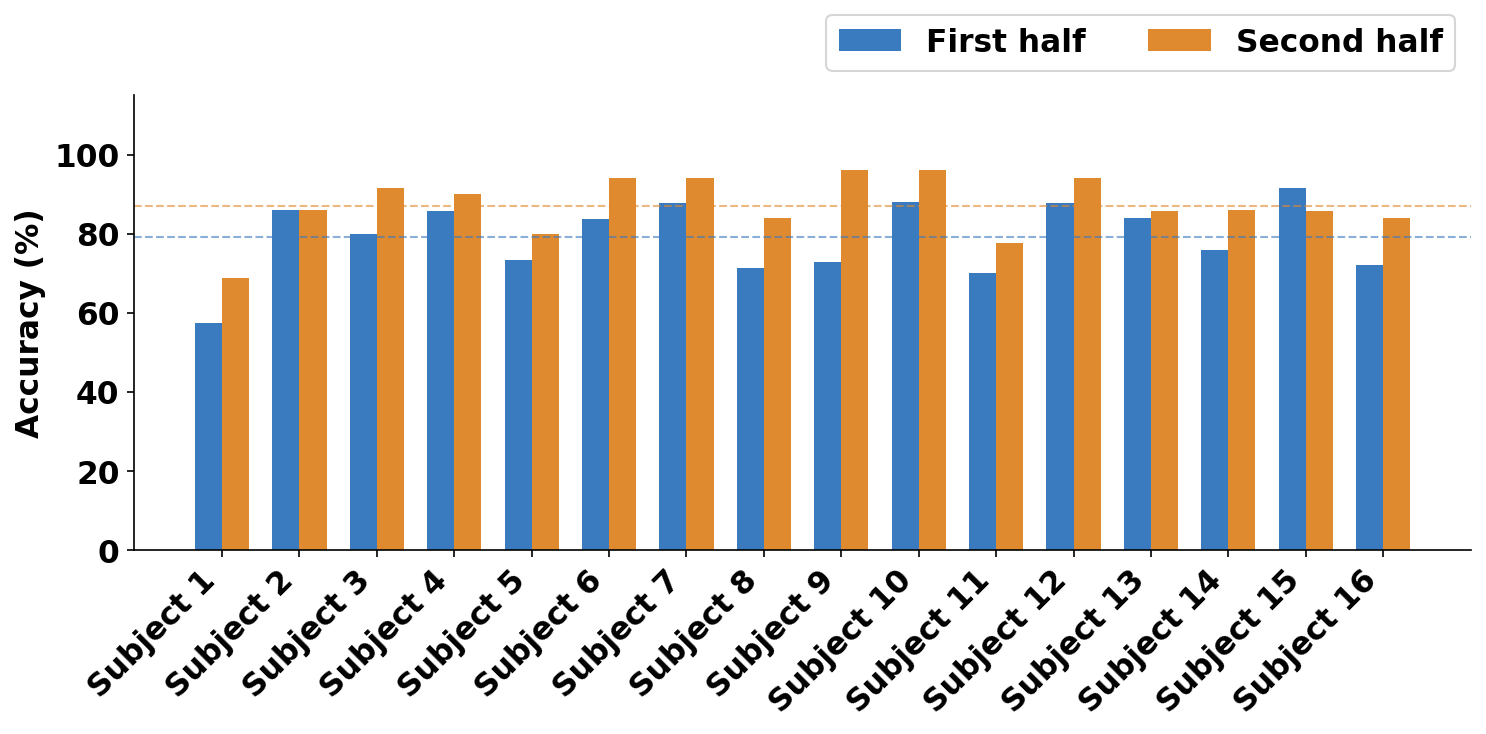}
        }
    \caption{Behavioral performance across participants based on comprehension-question responses. (a) Per-subject percentages of correct (blue), incorrect (amber), and \textit{``I could not pay attention"} (gray) responses; the dashed line indicates the mean accuracy. (b) Per-subject accuracy during the first and second experimental sessions, showing improved performance for most participants in the second session.}
    \label{fig:accuracy}
\end{figure*}

\subsubsection{Gaze}

Eye-tracking data (Fig.~\ref{fig:gaze_sample}) was collected at approximately 50 Hz using the \emph{Tobii Pro Glasses~3}. Measurements included three-dimensional gaze direction vectors, a two-dimensional scene-projected gaze coordinate, and pupil diameter for each eye. The two pupil channels are merged into one, taken from the right eye and falling back to the left where the right eye sample is missing, giving a 6-dimensional feature vector per sample. Approximately 15.1 hours of gaze data were recorded across all participants and trials. Missing samples were removed on a per-channel basis, linearly interpolated to a uniform 64 Hz grid, extended using the nearest valid values at the signal boundaries, and then low-pass filtered at 10 Hz using a fourth-order Butterworth filter.

\subsubsection{Egocentric video}

Scene video was recorded using the wide-angle camera integrated into the Tobii Pro Glasses 3, capturing the participant's first-person field of view at 25 fps and a resolution of $1920 \times 1080$ (H.264/AVC, $\approx$5.1 Mbps), along with a synchronized mono audio track sampled at 24 kHz. Each frame was downsampled to 160$\times$90 pixels and converted to grayscale. Dense optical flow was then computed between consecutive frames using the Farneback algorithm \cite{farneback2003two} and summarized by four features per frame: mean flow magnitude, standard deviation of flow magnitude, and the mean horizontal and vertical flow components. The resulting feature sequence was resampled from the frame rate to 64 Hz without further filtering, since the frame rate already bounds its bandwidth.

\subsubsection{IMU}

Head movement data was recorded at 120.6 Hz using the three-axis accelerometer and gyroscope integrated into the Tobii glasses, yielding a 6-dimensional feature (three-axis linear acceleration and three-axis angular velocity). Missing samples were removed prior to linear interpolation onto a uniform sampling grid at the median IMU sampling rate. The interpolated signal was resampled to 64 Hz and then low-pass filtered at 20 Hz with a fourth-order Butterworth filter.

\subsubsection{Audio}

Each audio stream was recorded as a stereo 16 kHz FLAC file: Device 1 for speakers $S_1$ and $S_2$, Device 2 for $S_3$ and $S_4$, and Device 3 for the two background-noise sources ($N_1$ and $N_2$). Because playback devices started at slightly different times (median offsets of approximately 80 ms and 160 ms relative to the earliest device), each waveform was aligned using its recorded playback timestamp before envelope extraction. 
Broadband amplitude envelopes were extracted using the Hilbert transform, low-pass filtered at 20 Hz with a fourth-order Butterworth filter, downsampled from 16 kHz to 64 Hz, and standardized per trial.

\section{Behavioral Data Analysis} \label{sec:behavioral}

\subsection{Comprehension Accuracy and Attentional Compliance}

Comprehension performance is summarized in Fig.~\ref{fig:accuracy}. Mean comprehension accuracy across all 16 participants was 83.2\%, with all but one participant (Subject~1, 63\%) achieving between 74\% and 92\% accuracy. Most participants showed improved performance in the second session, with the group mean increasing from 80\% to 87\%, suggesting a familiarization effect after the mandatory 10-minute break. Subject~1 showed the largest improvement (58\% to 69\%), while Subject~2 maintained the same accuracy (86\%) across both sessions.
The \textit{``I could not pay attention"} option was selected on only 1.4\% of trials, indicating that complete attentional failures were rare. Four participants (Subjects~2, 10, 14, and~16) never selected this option, while most others selected it on 1--3\% of trials. Subject~3 was an outlier, selecting it on 7\% of trials. Because these trials were excluded from the accuracy calculations, attentional non-compliance had minimal impact on the usable dataset. Correctness labels are released with the dataset, allowing researchers to include or exclude these trials in downstream analyses as desired.

\subsection{Eye and Head Movement Characteristics} \label{ssec:eye-head-char}

Mean gaze data quality was 80\%, measured as the average proportion of valid (non-missing) samples across trials and participants, with nine participants above 90\%. Tracking loss came primarily from eye blinks, partial occlusions, and gaze directions outside the camera's range during natural head and eye movements. Subjects 8 and 12 had substantially lower validity (27.3\% and 42.4\%), while the remaining participants averaged 86.5\%. Pupil diameter was bilaterally symmetric and within the normal range (left: $3.29 \pm 0.62$~mm; right: $3.28 \pm 0.57$~mm), and mean gaze position was near the center of the scene (horizontal: $0.50 \pm 0.11$; vertical: $0.56 \pm 0.18$) with a slight downward bias consistent with natural viewing.

\begin{figure}[t]
    \centering
    \includegraphics[width=\linewidth]{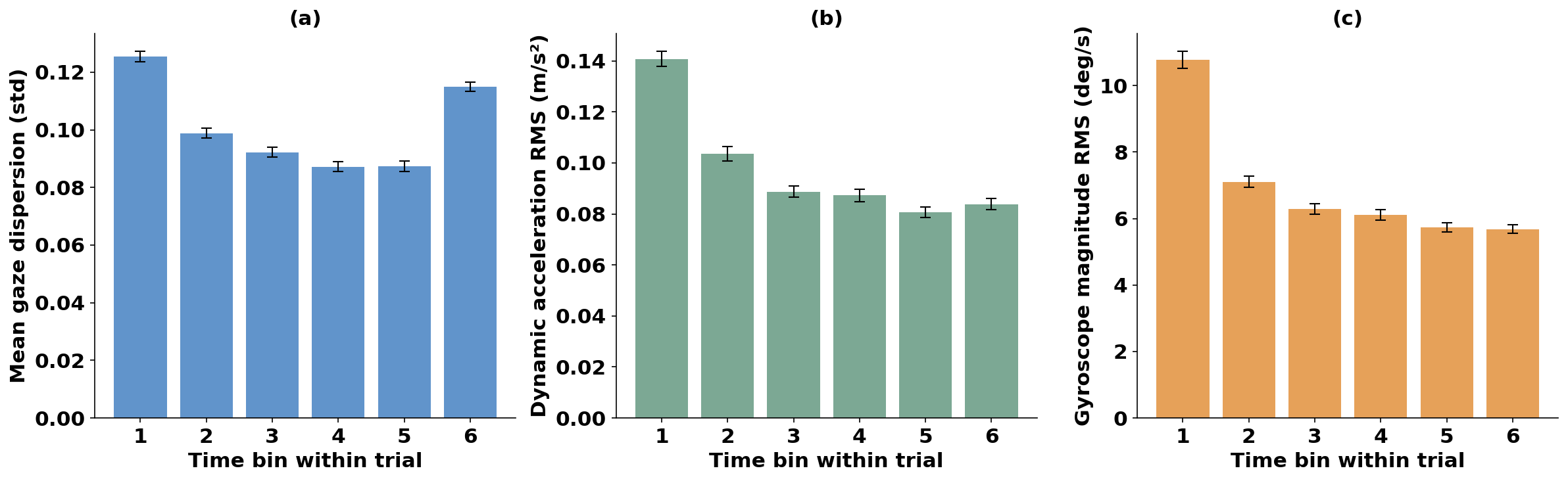}
    \caption{Mean gaze dispersion (a), dynamic acceleration RMS (b), and gyroscope magnitude RMS (c) across six equal time bins within a trial. Gaze dispersion peaks at trial onset and offset and is stable in between, indicating an initial orienting response followed by sustained attention. Head movement peaks at onset and stabilizes from Bin 3 onward with no rise at offset, indicating a postural adjustment followed by stillness during listening.}
    \label{fig:gaze_imu_char}
\end{figure}

Gaze dispersion was highest at trial onset and offset and stable in between (Fig.~\ref{fig:gaze_imu_char}), suggesting an initial orienting phase followed by sustained visual attention. Head movement followed the same pattern: the mean gyroscope RMS was 6.95 deg/s, greatest at onset and stable from Bin 3 onward. Subject 13 moved most (approximately 12 deg/s, over 1.5 times the group mean) and Subjects 3 and 11 least (3.6 and 4.0 deg/s).

\section{Evaluation Protocol and Baseline} \label{benchmark}

We evaluate MAESTRO on a four-speaker attention classification task that asks which of the four simultaneously presented speakers a listener is attending to (chance level of 25\%). We intentionally use a simple baseline to demonstrate that the dataset contains decodable signals and to provide a reference point for future work. The architecture follows an established AAD model and applies the same encoder design to every modality, so that configurations differ in which signals they receive rather than in how those signals are processed. The results should therefore be viewed as a lower bound on achievable performance rather than an upper bound.

\subsection{Training and Evaluation Splits}
We evaluate MAESTRO under within-subject and leave-one-subject-out (LOSO) settings. The within-subject setting measures performance when data from all participants are available for training, reflecting the most common protocol in the AAD literature; LOSO evaluates generalization to unseen individuals, a requirement for deployment where user-specific training data is unavailable. The official train and test splits are released with the dataset.

For the within-subject setting, data from all 16 participants are pooled and evaluated using 5-fold cross-validation. Folds are defined by trial content rather than participant-trial pairs: 80\% of the 100 unique trial contents (1,280 trials) are used for training and 20\% (320 trials) for testing, so each participant contributes to both partitions but never for the same content. For LOSO, models are trained on 15 participants and evaluated on the held-out participant, repeating for all 16. A fixed 20\% of trial contents is reserved for testing across all folds, so each held-out participant contributes a 20-trial test set against 1,200 training trials. In both settings, checkpoints are selected on a validation split drawn only from the training partition, and the held-out participant is never used for model selection.

Each 30-second trial is divided into one or more decision windows, which serve as model inputs. We evaluate window lengths of 5, 10, 15, 20, and 30 seconds, with a hop of half the window length for lengths below 30 seconds, giving eleven overlapping windows at 5\,s against a single window at 30\,s. Window length controls the temporal context available for each prediction and is applied consistently across both settings.

The model receives the four speech envelopes alongside the listener's signals, so any property that distinguishes the attended envelope from the others is a shortcut to the correct answer that bypasses the EEG. MAESTRO contains one. SNR adjustment scaled the attended channel before 16-bit PCM encoding, clipping its peaks but not the competitors'. Clipping flattens peaks, so the attended envelopes carry a lower crest factor (peak-to-RMS ratio) than the competing ones, 11.1\,dB against 20.5\,dB, and a logistic classifier trained on eight summary statistics of a standardized envelope identifies the attended speaker in 56\% of windows against a 25\% chance level.

Standardization does not remove this. Envelope extraction is linear and the standardization that follows is invariant to affine transformations, so it fixes only the mean and variance. Skewness, kurtosis, sparsity and dynamic range survive, because the distortion was not a change of scale. We therefore apply histogram equalization~\cite{heq1420370} across the four envelopes of each window: each envelope is sorted, the sorted values are averaged across the four, and each sample is replaced by the shared value at its own rank. All four then carry identical value distributions, so any statistic derived from amplitudes alone is identical by construction, and the probe falls to 26\%. What remains is the temporal ordering. Cortical tracking is driven by the timing of acoustic onset events rather than instantaneous amplitude~\cite{oganian2019speech}, so this is the information the decoder needs. The transform is not lossless and may remove some genuine envelope information, but it can only reduce decoding performance, never inflate it, so the accuracies reported here are conservative.

For each split, and window size, we assess whether every multimodal configuration statistically outperforms EEG alone using a two-tailed paired t-test.

\subsection{Baseline Network} \label{ssec:baseline}
\begin{figure}[t]
    \centering
    \includegraphics[width=\linewidth]{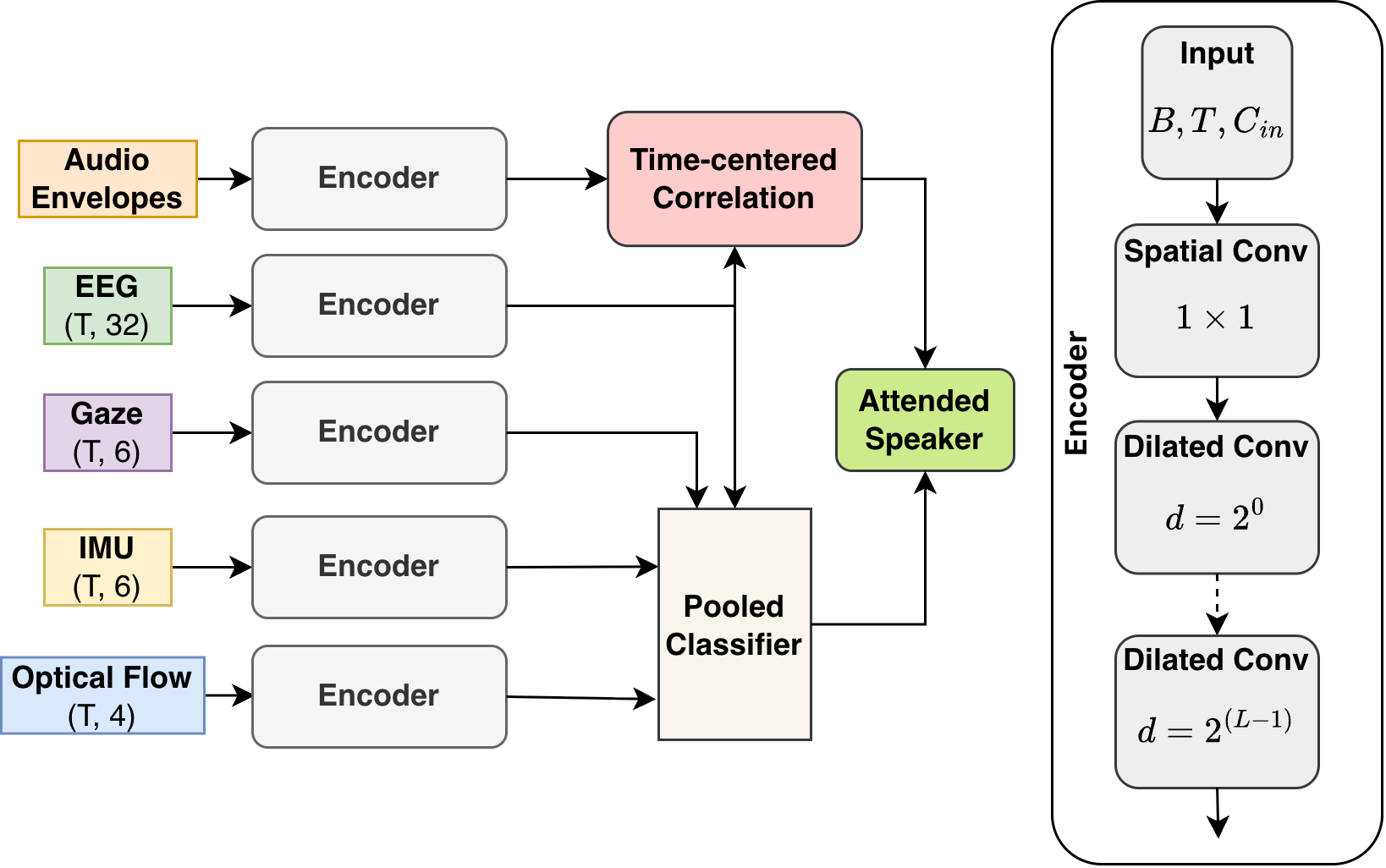}
    \caption{Baseline network for the benchmark task. Each modality is encoded by a dilated convolutional encoder (kernel size 3, dilation rates $2^0, \ldots, 2^4$, receptive field 0.98\,s) into a 16-dimensional embedding; the $1\times1$ spatial convolution in the encoder detail is applied to EEG only. The EEG embedding is compared with each speaker envelope by time-centered correlation, and every embedding is also pooled over time and classified over the four speaker positions with no audio input. The two scores are added, and the attended speaker is the one whose envelope occupies the highest-scoring slot.}
    \label{fig:baseline_classification}
\end{figure}

We adopt a multi-encoder architecture (Fig.~\ref{fig:baseline_classification}) based on the dilated convolutional model of Accou \textit{et al.}~\cite{accou2021modeling, accou2021predicting}, which encodes EEG and audio signals using dilated convolutional encoders and identifies the attended speaker through similarity between the resulting embeddings. We retain this framework for EEG and extend it to multimodal decoding with additional encoders for gaze, IMU, and scene video.

\balance

All encoders use a kernel size of 3 and produce 16-dimensional embeddings. Each uses 5 layers with dilation rates $2^0, 2^1, \ldots, 2^4$, giving a receptive field of 63 samples (0.98\,s at 64\,Hz), which comfortably spans the 0-400\,ms lags over which a cortical response to a speech envelope unfolds~\cite{osullivan2015, crosse2016multivariate}. Encoder architectures are identical across all five window sizes, rather than retuned per window, to isolate the effect of window size from architecture changes. The convolutions are centered rather than causal, reflecting that the cortical response to audio at time $t$ appears in the EEG after a short delay. The EEG encoder additionally includes a $1\times1$ spatial convolution with 8 filters before the dilated stack to mix information across the 32 EEG channels. Group normalization follows each convolution. A rectifier follows every layer except the last, which is linear so that embeddings can take negative values. The audio envelope encoder uses one input channel and shares weights across all four envelope streams.

EEG representations are compared with the four speaker envelope representations through correlation over time, giving one score per slot (i.e., speaker index). Gaze, IMU, and video carry a low temporal relationship to a speech envelope, and instead indicate which speaker the listener is facing; these are classified over the four speaker positions without any audio input, giving a second score per slot. Let $\mathbf{z}$ denote the EEG embedding over the $T$ samples of the decision window and $\hat{\mathbf{a}}_k$ the embedding of the envelope in slot $k$. Let $\mathbf{z}_d$ denote the $d$-th dimension of the embedding and $\overline{\mathbf{z}}_d$ its mean over time. The score $s_k$ for slot $k$ is obtained from the 16 per-dimension correlations $c_d(k)$:
\begin{align}
c_d(k) &= \frac{\big\langle \mathbf{z}_d - \overline{\mathbf{z}}_d,\; \hat{\mathbf{a}}_{k,d} - \overline{\hat{\mathbf{a}}}_{k,d} \big\rangle}{\big\lVert \mathbf{z}_d - \overline{\mathbf{z}}_d \big\rVert_2 \; \big\lVert \hat{\mathbf{a}}_{k,d} - \overline{\hat{\mathbf{a}}}_{k,d} \big\rVert_2}, \notag \\[2pt]
s_k &= \tau \sum_d w_d\, c_d(k) \;-\; \frac{1}{4}\sum_{j=1}^{4} \tau \sum_d w_d\, c_d(j),
\label{eq:coupling}
\end{align}
where $c_d(k)$ is the Pearson correlation over time between dimension $d$ of the two embeddings, $\mathbf{w}$ is a learned weight vector combining the 16 correlations, and $\tau$ is a learned temperature. Both signals are centered in time before the correlation is taken. Without centering, an encoder that ignored the EEG and emitted a constant embedding would still yield unequal scores, set by the speaker envelopes alone. With centering that embedding gives $\mathbf{z} - \overline{\mathbf{z}} = 0$, so every correlation is zero, the four scores are equal, and accuracy falls to 25\%. Correlation also normalizes by magnitude, so an envelope cannot be favored by being larger.

The second score is computed from pooled statistics. Each modality embedding is reduced to its mean and temporal standard deviation over the window and passed through a two-layer perceptron. EEG contributes here as well, since attending to one side produces lateralized cortical activity. The per-modality embeddings are concatenated and passed through a fusion head of the same form, whose logits are defined over the four speaker positions and mapped into slot order before being added to the correlation scores. The prediction is the highest-scoring slot, and the attended speaker is the one whose envelope occupied it.

We evaluate all 15 input configurations supported by MAESTRO. The modalities are fused within a single model trained end-to-end, rather than by averaging the predictions of separately trained single-modality models. To keep any one modality from dominating, we apply modality dropout: at each training step, each modality is withheld independently with probability 0.3, and never all at once. It is disabled at evaluation and not applied to single-modality configurations. Each model is trained end-to-end with the AdamW optimizer (learning rate $10^{-3}$, weight decay $10^{-4}$) and gradient clipping (maximum norm 1.0), on the objective

\begin{equation}
\begin{split}
\mathcal{L} = {}& \mathcal{L}_{\mathrm{CE}} + 0.3\,\mathcal{L}_{\mathrm{aux}} + 1.0\,\mathcal{L}_{\mathrm{con}} \\
& + 0.5\,\mathcal{L}_{\mathrm{hinge}} + 0.1\,\mathcal{L}_{\mathrm{coll}} + 0.3\,\mathcal{L}_{\mathrm{adv}},
\end{split}
\label{eq:loss}
\end{equation}

\begin{table*}[t]
\centering
\caption{Attended-speaker accuracy (mean $\pm$ std across folds, \%) for the within-subject and leave-one-subject-out (LOSO) splits, across all 15 modality combinations and 5 window sizes. Bold marks the best-performing mode within each window-size column, computed separately per split. $^*$ marks a configuration that significantly outperforms EEG alone ($p < 0.05$, two-tailed paired $t$-test).}
\label{tab:aad_results}
\resizebox{\textwidth}{!}{%
\begin{tabular}{rccccccccccc}
\toprule
 & \multicolumn{5}{c}{Within-subject} & \multicolumn{5}{c}{LOSO} \\
\cmidrule(lr){2-6} \cmidrule(lr){7-11}
Mode & 5s & 10s & 15s & 20s & 30s & 5s & 10s & 15s & 20s & 30s \\
\midrule
EEG & 43.63$\pm$1.76 & 47.57$\pm$4.08 & 53.17$\pm$1.66 & 53.62$\pm$2.67 & 59.06$\pm$2.04 & 43.64$\pm$3.18 & 50.19$\pm$6.98 & 52.40$\pm$7.02 & 55.31$\pm$9.92 & 61.88$\pm$11.71 \\
Gaze & 37.13$\pm$1.16 & 37.58$\pm$1.75 & 39.98$\pm$2.10 & 37.95$\pm$2.38 & 36.04$\pm$2.79 & 33.99$\pm$8.05 & 34.95$\pm$9.38 & 36.08$\pm$9.09 & 36.87$\pm$8.98 & 35.46$\pm$9.44 \\
IMU & 38.21$\pm$2.80 & 38.75$\pm$2.13 & 38.84$\pm$2.44 & 34.85$\pm$4.16 & 35.41$\pm$4.14 & 37.07$\pm$10.86 & 34.77$\pm$9.46 & 38.48$\pm$13.53 & 32.16$\pm$11.29 & 35.46$\pm$12.69 \\
Video & 43.71$\pm$4.20 & 42.59$\pm$4.90 & 44.33$\pm$3.86 & 43.16$\pm$2.44 & 43.37$\pm$6.63 & 39.57$\pm$12.71 & 43.56$\pm$14.66 & 41.35$\pm$13.65 & 41.56$\pm$17.61 & 44.69$\pm$14.52 \\
\midrule
EEG+Gaze & 51.24$\pm$3.89$^*$ & 55.27$\pm$1.79$^*$ & 57.97$\pm$1.61$^*$ & 59.73$\pm$3.41$^*$ & 65.15$\pm$3.83 & 50.54$\pm$7.52$^*$ & 52.14$\pm$10.16 & 55.90$\pm$8.64 & 61.00$\pm$10.49$^*$ & 62.45$\pm$16.23 \\
EEG+IMU & 49.14$\pm$2.50$^*$ & 51.50$\pm$3.97 & 54.13$\pm$4.12 & 57.91$\pm$3.42 & 64.08$\pm$2.67$^*$ & 48.54$\pm$6.31$^*$ & 50.93$\pm$7.82 & 55.80$\pm$8.66 & 59.28$\pm$8.57$^*$ & 63.04$\pm$13.38 \\
EEG+Video & 49.90$\pm$4.78$^*$ & 52.70$\pm$2.68 & 54.69$\pm$2.01 & 59.53$\pm$2.74$^*$ & 62.56$\pm$2.20$^*$ & 50.88$\pm$7.44$^*$ & 54.19$\pm$6.69 & 56.25$\pm$8.41 & 60.94$\pm$10.11$^*$ & 63.44$\pm$12.46 \\
Gaze+IMU & 45.34$\pm$2.90 & 43.51$\pm$2.46 & 43.28$\pm$3.25 & 39.89$\pm$2.68 & 38.30$\pm$2.14 & 40.53$\pm$14.05 & 41.24$\pm$9.15 & 40.30$\pm$9.70 & 34.98$\pm$7.42 & 34.84$\pm$8.58 \\
Gaze+Video & 48.89$\pm$2.98$^*$ & 49.76$\pm$3.86 & 49.64$\pm$4.06 & 50.94$\pm$3.64 & 52.81$\pm$3.14 & 46.87$\pm$16.23 & 47.10$\pm$14.10 & 46.78$\pm$13.14 & 45.50$\pm$17.99 & 44.61$\pm$15.70 \\
IMU+Video & 42.93$\pm$4.34 & 43.94$\pm$4.05 & 45.15$\pm$4.07 & 43.98$\pm$4.16 & 46.24$\pm$3.82 & 41.55$\pm$12.21 & 42.33$\pm$15.15 & 40.77$\pm$14.23 & 43.60$\pm$14.45 & 44.54$\pm$16.36 \\
\midrule
EEG+Gaze+IMU & 52.22$\pm$2.42$^*$ & 54.93$\pm$1.46$^*$ & 59.62$\pm$2.30$^*$ & 61.33$\pm$1.98$^*$ & 66.96$\pm$1.00$^*$ & 51.23$\pm$10.65$^*$ & 54.78$\pm$11.84 & 57.28$\pm$10.35 & 59.73$\pm$10.80$^*$ & 62.70$\pm$15.19 \\
EEG+Gaze+Video & 55.57$\pm$1.84$^*$ & \textbf{58.45$\pm$3.86}$^*$ & 59.61$\pm$1.84$^*$ & 62.11$\pm$3.98$^*$ & \textbf{69.83$\pm$2.33}$^*$ & 53.47$\pm$12.28$^*$ & \textbf{57.33$\pm$10.37}$^*$ & 59.57$\pm$11.41$^*$ & 60.98$\pm$11.18$^*$ & 63.98$\pm$13.31 \\
EEG+IMU+Video & 52.86$\pm$4.74$^*$ & 54.56$\pm$5.42 & 57.03$\pm$3.57$^*$ & 60.60$\pm$2.61$^*$ & 62.38$\pm$6.40 & 50.34$\pm$9.03$^*$ & 55.94$\pm$9.73 & \textbf{59.59$\pm$11.15}$^*$ & \textbf{61.29$\pm$9.32}$^*$ & 64.31$\pm$14.06 \\
Gaze+IMU+Video & 48.88$\pm$3.03$^*$ & 50.51$\pm$3.75 & 50.89$\pm$3.14 & 52.88$\pm$3.10 & 54.13$\pm$3.35 & 44.77$\pm$16.86 & 45.76$\pm$17.00 & 46.77$\pm$16.80 & 44.59$\pm$16.39 & 44.59$\pm$16.39 \\
\midrule
EEG+Gaze+IMU+Video & \textbf{56.64$\pm$3.28}$^*$ & 58.01$\pm$1.97$^*$ & \textbf{61.20$\pm$1.83}$^*$ & \textbf{62.80$\pm$2.72}$^*$ & 67.08$\pm$2.03$^*$ & \textbf{54.41$\pm$11.83}$^*$ & 55.87$\pm$11.51 & 57.28$\pm$11.20 & 59.75$\pm$10.08 & \textbf{67.45$\pm$18.58} \\
\bottomrule
\end{tabular}}
\end{table*}

where $\mathcal{L}_{\mathrm{CE}}$ is the cross-entropy over the four slot scores with label smoothing (0.1). Writing $B$ for the batch size, $M$ for the number of modalities present, $D=16$ for the embedding width, $y$ for the slot holding the attended envelope, and $\zeta$ for the softplus, the remaining terms are-
\begin{align}
\mathcal{L}_{\mathrm{aux}} &= \tfrac{1}{M}\textstyle\sum_{m}\mathrm{CE}\big(\mathbf{g}^{(m)}, y\big), \\
\mathcal{L}_{\mathrm{con}} &= -\tfrac{1}{2B}\textstyle\sum_{i}\Big[\log\tfrac{e^{S_{ii}}}{\sum_j e^{S_{ij}}} + \log\tfrac{e^{S_{ii}}}{\sum_j e^{S_{ji}}}\Big], \\
\mathcal{L}_{\mathrm{hinge}} &= \tfrac{1}{B}\textstyle\sum_{i}\big[\zeta(\tilde{s}_i - s_i + \delta) + \zeta(\bar{s}_i - s_i + \delta)\big], \\
\mathcal{L}_{\mathrm{coll}} &= \tfrac{1}{BD}\textstyle\sum_{i,d}\max(0,\gamma-\sigma_{i,d}) + \tfrac{\lambda}{D}\textstyle\sum_{d\neq d'} C_{dd'}^{2}, \\
\mathcal{L}_{\mathrm{adv}} &= \mathrm{CE}\big(\mathbf{h}(\hat{\mathbf{a}}_1,\ldots,\hat{\mathbf{a}}_4), y\big).
\end{align}
$\mathcal{L}_{\mathrm{aux}}$ applies the speaker-position loss to each modality's logits $\mathbf{g}^{(m)}$ separately, so the strongest modality does not absorb the gradient. $\mathcal{L}_{\mathrm{con}}$ is a contrastive term over the batch, where $S_{ij}$ scores window $i$'s EEG against window $j$'s attended envelope and the correct pairing must win; the batch is drawn from one participant, so the pairing cannot be made from listener identity. $\mathcal{L}_{\mathrm{hinge}}$ requires the score $s_i$ from the real signals to beat the score $\tilde{s}_i$ from another window's and $\bar{s}_i$ from zeros by a margin $\delta=0.5$. $\mathcal{L}_{\mathrm{coll}}$ requires the temporal standard deviation $\sigma_{i,d}$ of each embedding dimension to reach $\gamma=0.5$ and penalizes the off-diagonal covariance $C$ between dimensions ($\lambda=0.04$), since a constant embedding carries no information. $\mathcal{L}_{\mathrm{adv}}$ is a classifier $\mathbf{h}$ that reads only the audio embeddings, trained through a gradient-reversal layer so the audio encoder unlearns any cue that identifies the attended speaker.

Batches contain 32 windows and are drawn from one participant at a time. The learning rate is halved after five epochs without validation improvement, with a minimum value of $10^{-6}$, and training stops after twelve epochs without improvement, up to a maximum of 50 epochs.

\section{Benchmark Results} \label{results}

\subsection{Within-subject}

We report results for all single-modality and multimodal input configurations supported by MAESTRO (Table~\ref{tab:aad_results} (left)). Reported means and standard deviations are computed across 5 cross-validation folds, with each fold trained once. For every window size, the best-performing multimodal configuration outperforms EEG alone in the within-subject setting, with gains ranging from 8.03\% (15\,s) to 13.01\% (5\,s). EEG is also the strongest single modality at every window size except 5\,s, so these gains arise from complementary information across modalities rather than from replacing a weaker modality with a stronger one.
The optimal modality combination varies with window size: all four modalities perform best at 5\,s, 15\,s, and 20\,s, and EEG+gaze+video at 10\,s and 30\,s. Notably, the full four-modality configuration is optimal for three of the five window sizes and every winning combination contains EEG, suggesting that visual and motor signals act as complements to the neural signal rather than as substitutes for it. Each of these best-performing configurations significantly outperforms EEG alone at every window size.

\begin{figure*}[t]
    \centering
    \includegraphics[width=\textwidth]{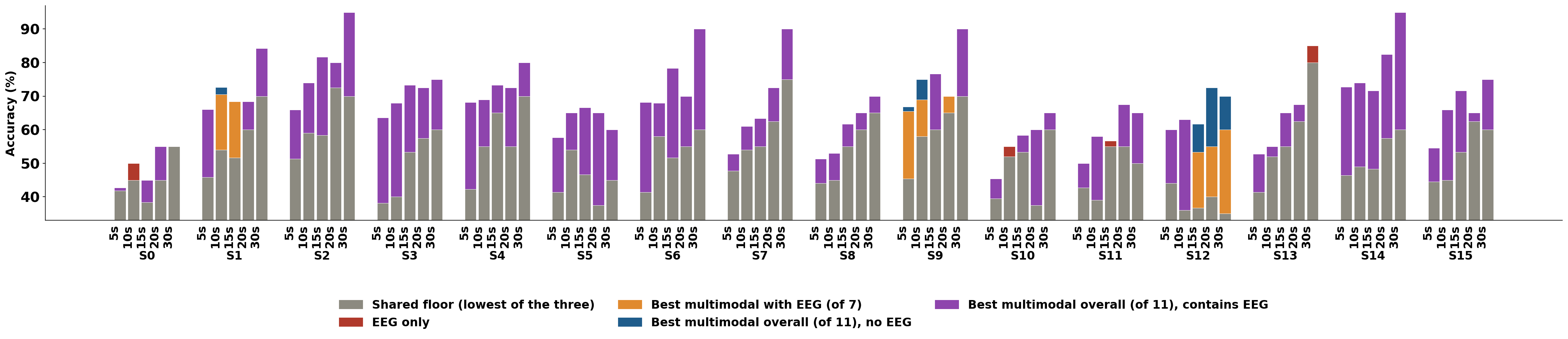}
    \caption{Per-subject and per-window comparison under LOSO evaluation of EEG alone, the best of the seven multimodal combinations containing EEG, and the best of all eleven; the second set is a subset of the third, so the two often coincide. Each subject is shown as five stacked bars, one per window size (5, 10, 15, 20, 30\,s). Gray marks the lowest of the three accuracies, and the segments above it are colored by which configuration attained each: red (EEG alone), orange (best with EEG), blue (best overall, no EEG), purple (best overall, contains EEG).}
    \label{fig:aad_loso_bar}
\end{figure*}

\subsection{Leave-One-Subject-Out} \label{loso} 

We report mean accuracy across subjects for all single-modality and multimodal input configurations (Table~\ref{tab:aad_results} (right)). Reported means and standard deviations are computed across the 16 held-out subjects, with each subject trained once. Under LOSO evaluation, multimodal fusion improves performance at every window size, with gains over EEG alone ranging from 5.58\% (30\,s) to 10.77\% (5\,s). EEG is itself the strongest single modality throughout, so these margins also describe the improvement over the best unimodal baseline. They are nevertheless narrower than their within-subject counterparts (8.03--13.01\%) and the across-subject standard deviations are roughly four times larger, indicating that generalization to unseen subjects remains a more challenging problem. As in the within-subject results, no single modality combination dominates: all four modalities perform best at 5\,s, EEG+gaze+video at 10\,s, EEG+IMU+video at 15\,s and 20\,s, and all four again at 30\,s. These configurations significantly outperform EEG alone at four of the five window sizes, the exception being 30\,s, where only twenty test windows per subject leave the test underpowered.

These averages, however, do not capture subject-level variability. Fig.~\ref{fig:aad_loso_bar} compares EEG-only performance against each subject's best-performing EEG-containing and best-performing overall multimodal configurations. Across all 80 subject-window combinations, the best overall configuration outperforms EEG only in 75 cases (93.8\%, mean gain 15.66\% among wins), ties in one, and trails in only 4 (5.0\%, mean deficit 3.67\%). Restricting to EEG-containing configurations yields the same win, tie, and loss counts, with a mean gain of 15.05\%, and the best EEG-containing multimodal configuration is itself the overall best in 72 of 80 cases (90.0\%). This suggests multimodal inputs benefit nearly all subjects, and that the benefit is carried by combinations retaining the neural signal rather than by behavioral signals alone.

\subsection{SNR-Stratified Analysis}

\begin{figure}[b]
    \centering
    \includegraphics[width=0.8\linewidth]{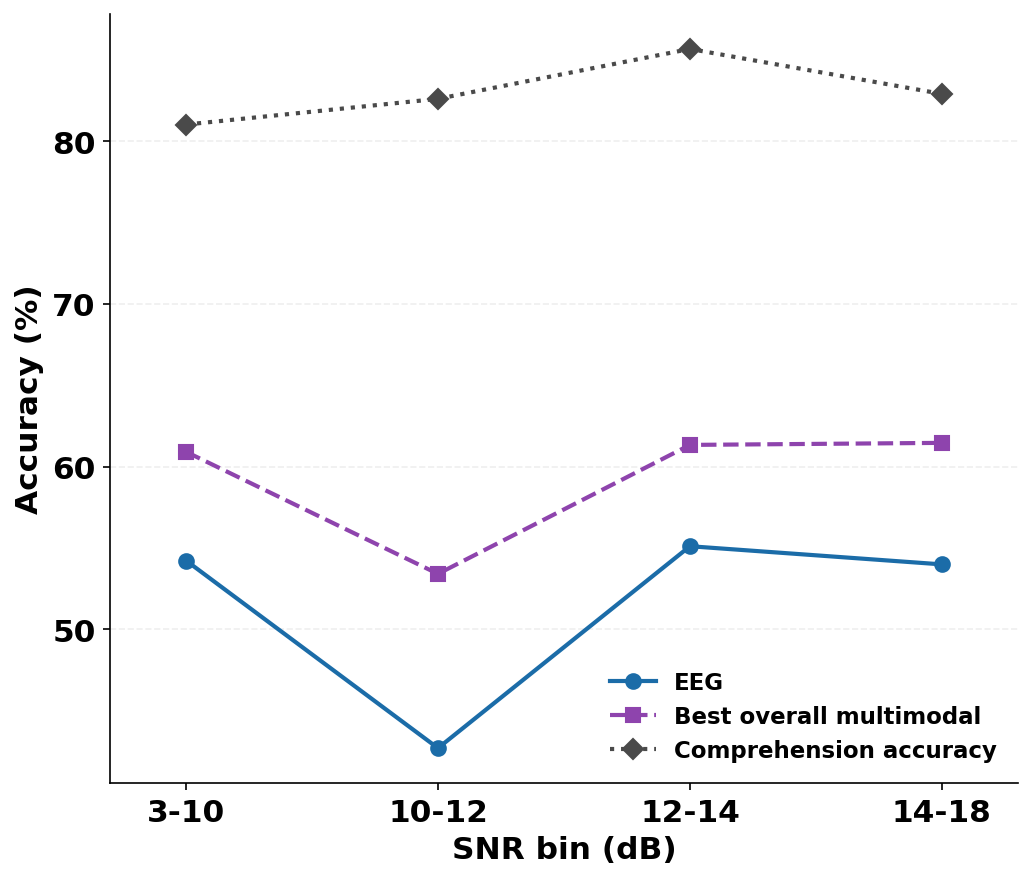}
    \caption{Attended-speaker accuracy versus SNR bin under the LOSO protocol, averaged across all five window sizes. Lines show EEG alone, and the best-performing multimodal configuration overall, alongside comprehension-question accuracy averaged across subjects per SNR bin.}
    \label{fig:snr_str}
\end{figure}

To characterize how decoding performance under LOSO evaluation depends on listening difficulty, we evaluated LOSO accuracy as a function of SNR, averaged across all five window sizes, as shown in Fig.~\ref{fig:snr_str}. Because SNRs in MAESTRO are approximately normally distributed, we partitioned trials into four SNR bins (3-10, 10-12, 12-14, and 14-18 dB) rather than grouping by individual SNR value, which would yield uneven sample counts and unreliable tail estimates; bins were computed once from the full dataset and applied consistently across window sizes. For each window size, we evaluate EEG alone and the best-performing multimodal combination, selected independently of the SNR-bin breakdown, and average accuracy across window sizes per bin. Accuracy is evaluated using the fold-specific model on held-out windows per bin, averaged across the sixteen held-out subjects and then across window sizes; comprehension-question accuracy is shown alongside for reference.

Decoding accuracy showed little systematic dependence on SNR. EEG-only accuracy was 54.2\% in the lowest bin and 54.0\% in the highest, and the best-performing multimodal configuration moved from 60.9\% to 61.5\%, remaining above EEG in every bin by 6.2 to 10.7 points. Every winning combination contains EEG. This holds at each window size and within each SNR bin, so no purely behavioral combination is best at any SNR level. Both curves instead share a pronounced dip in the 10-12 dB bin, to 42.7\% and 53.4\% respectively, that recurs at all five window sizes. The bin edges are quantiles of the full dataset, but the held-out test contents do not divide evenly among them: this bin holds 3 of the 20 test trials per listener against 5, 6 and 6 for the others, so all four attended-speaker positions cannot appear in it. Every LOSO fold is tested on the same 20 contents, so the same three trials fall in this bin for every listener, and the 16 fold accuracies are repeated measurements of those three trials rather than independent samples of the 10-12\,dB condition. We cannot rule out that this SNR range is genuinely harder, but comprehension accuracy is lowest in the 3-10\,dB bin rather than in the 10-12\,dB bin, so listeners did not find this range hardest. Comprehension accuracy likewise varied little across bins (81.0\% to 82.6-85.7\%). These results are consistent with prior work reporting decoding accuracy relatively insensitive to SNR despite degrading neural tracking strength \cite{wang2020robust}.

\begin{figure}[t]
    \centering
    \includegraphics[width=0.85\linewidth]{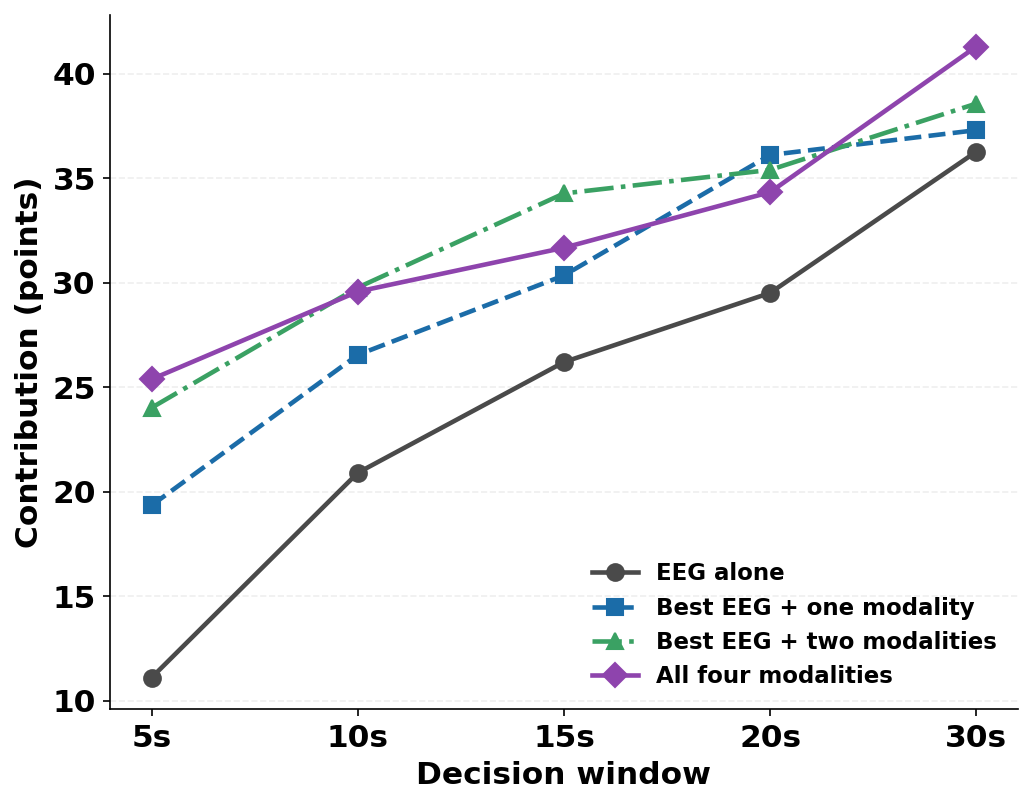}
    \caption{Contribution of the EEG-containing configurations under LOSO evaluation, defined as accuracy minus the accuracy the same model reaches when the listener's signals are permuted across test windows. Curves show EEG alone, the best EEG-containing pair and triple, and the full four-modality configuration; the pair and triple are selected by contribution at each window size, so their identity varies across the sweep.}
    \label{fig:contribution_loso}
\end{figure}

\subsection{Attributing Accuracy to the Recorded Signals} \label{ssec:null}

A decoder that found any residual cue in the speech envelopes could reach a high accuracy without using the listener's signals at all. We therefore test whether the accuracies reported above depend on those signals by permuting them across test windows, so that each window receives another window's EEG, gaze, IMU, and video while keeping its own envelopes and its own label. 
A single permutation is applied to all modalities together, so their correspondence within a window is preserved. The difference between the real and the permuted accuracy is the \emph{contribution}, the part of an accuracy attributable to the listener's signals. We average over 20 permutations. Fig.~\ref{fig:contribution_loso} reports the contribution of the EEG-containing configurations under LOSO. Every configuration has a positive contribution at every window size, from $+11.1$ points for EEG alone at 5\,s to $+41.3$ points for the full configuration at 30\,s. EEG alone has the lowest contribution at every window, and the gap is widest at 5\,s, where adding the three behavioral modalities raises the contribution from $+11.1$ to $+25.4$ points. The behavioral modalities therefore increase not only accuracy but the share of it that depends on the listener's signals. The multimodal advantage reported above therefore reflects better use of the listener's signals, not better use of the speech envelopes.

\section{Discussion} \label{discussion}

\subsection{Interpretation of Multimodal Benefit} \label{sec:interpretation}
Incorporating behavioral modalities alongside EEG improves decoding at every window size and under both splits, by 8.03 to 13.01\% within-subject and 5.58 to 10.77\% under LOSO. The two splits should be read separately rather than compared cell by cell: within-subject folds rotate through all 100 trial contents, whereas every LOSO fold is evaluated on the same held-out 20, so the two halves of Table \ref{tab:aad_results} result from different test material. The optimal combination varies with window size rather than converging on a single configuration, but every winning combination contains EEG, and no purely behavioral combination is best at any window size. This pattern is broadly consistent with the underlying signals: EEG carries the envelope-tracking information the task requires, while gaze, head motion, and scene video indicate where the listener is oriented and narrow the four-way choice. The benefit is largest at short windows, where EEG has least context to work with: under LOSO the gain nearly doubles as the window shortens, from 5.58\% at 30\,s to 10.77\% at 5\,s. The permutation analysis of Section~\ref{ssec:null} shows that this benefit is not an artifact of the speech envelopes: adding the behavioral modalities raises the contribution as well as the accuracy. Subject-level analyses further show that each participant's best-performing multimodal configuration outperforms EEG alone in most subject-window combinations (Section~\ref{loso}), indicating that the observed gains are broad and consistent rather than artifacts of population averaging.
Together, the results suggest that neural and behavioral signals provide complementary information, and integrating them can improve decoding across future hearing-assistive technologies.

\subsection{Relation to Prior Work}
EEG alone reaches 43.63\%-59.06\% across window sizes in the within-subject setting, well below the 80-99\% reported for binary two-speaker decoding~\cite{neurotoken2026}. Part of the gap is task difficulty: four competing speakers, varying SNRs, and background noise lower the prior probability of each class and complicate the acoustic scene. The rest reflects evaluation, since prior benchmarks do not prevent data leakage and such results largely fall to chance under a strict regime. EEG alone still recovers the attended speaker at more than twice chance at the longest window, and the full multimodal configuration reaches 67.45\% under LOSO. The behavioral modalities decode the attended speaker well above chance without any access to the speech: scene video reaches 39.57\%-44.69\% under LOSO and gaze with video 44.61\%-47.10\%. This is consistent with prior findings that auditory attention is reflected in overt behavior as well as neural activity. Hendrikse \textit{et al.}~\cite{hendrikse2019} showed that listeners naturally orient their gaze toward attended speakers, and Rotaru \textit{et al.}~\cite{rotaru2024} that gaze-related signals may confound EEG-only decoders. None of these modalities matches EEG on its own, but each adds several points when combined with it. Separating neural from behavioral contributions therefore requires datasets that record unconstrained gaze and head movement rather than suppressing them.

\subsection{Implications for Hearing Aid Design}
The benchmark results have implications for neuro-steered hearing aid design. Eye-tracking and head-mounted inertial sensors are already present in wearable form factors, and the results here indicate that they carry attention-related information that complements EEG, particularly at shorter decision windows where EEG has least context to work with (Section~\ref{sec:interpretation}). Because the best-performing combination varies with window size and across subjects, practical systems will likely benefit from adaptive, user- and context-dependent modality weighting rather than a fixed sensor configuration. Furthermore, MAESTRO's systematic SNR variation provides a platform for studying how such systems respond to increasingly challenging acoustic conditions, a key requirement for real-world hearing-aid deployment \cite{smeds2015estimation}.

\subsection{Limitations and Future Directions}
MAESTRO has several limitations. Its 16 participants are comparable to established AAD datasets such as KUL~\cite{biesmans2016}, but the small cohort limits cross-subject analyses and increases sensitivity to outliers such as Subject 1, whose comprehension accuracy (63\%) suggests inconsistent attentional compliance. The dataset is also limited to native English speakers, restricting applicability to other languages, and gaze quality varied across participants, with Subjects 8 and 12 showing substantially lower tracking validity, likely from equipment fitting; per-subject quality statistics are released with the dataset. The acoustic scene is fixed: the four loudspeakers occupy the same positions throughout, so orienting toward the attended speaker reduces to a choice among four known directions, and gaze dispersion stabilizes shortly after trial onset. The behavioral modalities therefore face a simpler problem than in a realistic environment, where sources move and their number is unknown in advance. Repeating the paradigm across rooms, with moving or repositioned sources, would establish how far the multimodal benefit transfers.

The released audio also carries a rendering artifact. The SNR adjustment clipped the attended channel but not the competitors, flattening its waveform (crest factor 11.1\,dB against 20.5\,dB). Our benchmark removes the resulting envelope cue, but the distortion is a property of the released recordings and affects any downstream use of the raw audio.

The baseline applies the same encoder design to all modalities despite their differing temporal and noise characteristics, with only the EEG encoder receiving a spatial mixing layer. Future work should explore modality-specific encoders and fusion that weights modalities by signal quality, and the fixed modality dropout rate could be made adaptive.

MAESTRO is also well suited for attended-speaker extraction, where the goal is to isolate the target rather than identify it. Recent EEG-guided extraction systems have been evaluated only in two-speaker scenarios with EEG as the sole attention signal~\cite{pan10683957, pan2024neuroheed+}; MAESTRO's egocentric video, gaze, head motion, and naturally mixed audio provide a foundation for extending them to four-speaker environments. The dataset also enables direct investigation of the gaze-confound hypothesis of Rotaru \textit{et al.}~\cite{rotaru2024}, and its combined gaze, head-motion, EEG, pupillometry, and SNR annotations support studies of listening effort~\cite{dimitrijevic2019neural, seifi2020exploratory}, with potential applications in hearing-aid fitting, fatigue monitoring, and personalized hearing assistance.

\section{Conclusion} \label{conclusion}
We release MAESTRO, a multimodal AAD dataset containing synchronized recordings from 16 participants across 1,600 trials, together with a reproducible evaluation framework spanning a four-speaker attention decoding benchmark, five decision window sizes, and both within-subject and leave-one-subject-out evaluations. Auditory attention decoding has traditionally been viewed as a purely neural problem. MAESTRO reframes it as a multimodal sensing problem in which EEG, gaze, head motion, and scene video each provide complementary information about attention. As the first AAD dataset to simultaneously capture these modalities under naturalistic conditions with unconstrained participant behavior, MAESTRO enables investigations that were previously impossible.
Our benchmarks demonstrate that behavioral signals are not merely auxiliary to EEG but encode substantial attention-related information in their own right, identifying the attended speaker well above chance with no access to the speech. Adding them to EEG improves decoding at every window size and under both splits, and the best combination varies with window size, which motivates adaptive rather than fixed modality weighting.

We hope MAESTRO shifts the focus from simply improving EEG decoding accuracy to understanding what is being decoded, how different modalities encode attentional information, and how they can be combined to build more robust and interpretable auditory attention decoders.

\section{Acknowledgement}

We gratefully acknowledge computational resources provided by the Ohio Supercomputer Center (OSC) and support from the National Science Foundation (NSF) (IIS-2235228).

\bibliographystyle{IEEEtran}
\bibliography{reference}

@article{folstein1975mini,
  title     = {`{M}ini-mental state': A practical method for grading 
               the cognitive state of patients for the clinician},
  author    = {Folstein, Marshal F. and Folstein, Susan E. and McHugh, Paul R.},
  journal   = {Journal of Psychiatric Research},
  volume    = {12},
  number    = {3},
  pages     = {189--198},
  year      = {1975},
  publisher = {Elsevier},
  doi       = {10.1016/0022-3956(75)90026-6}
}

@misc{resound_hearing_test,
  author       = {{ReSound}},
  title        = {Free Online Hearing Test},
  howpublished = {\url{https://www.resound.com/en-us/online-hearing-test}},
  note         = {Accessed: Sep. 6, 2026}
}

@inproceedings{panayotov2015librispeech,
  title={Librispeech: an asr corpus based on public domain audio books},
  author={Panayotov, Vassil and Chen, Guoguo and Povey, Daniel and Khudanpur, Sanjeev},
  booktitle={IEEE International Conference on Acoustics, Speech and Signal Processing (ICASSP)},
  pages={5206--5210},
  year={2015},
  organization={IEEE}
}

@inproceedings{foster2015chime,
  title={Chime-home: A dataset for sound source recognition in a domestic environment},
  author={Foster, Peter and Sigtia, Siddharth and Krstulovic, Sacha and Barker, Jon and Plumbley, Mark D},
  booktitle={IEEE Workshop on Applications of Signal Processing to Audio and Acoustics (WASPAA)},
  pages={1--5},
  year={2015},
  organization={IEEE}
}

@article{cherry1953,
  author  = {Cherry, E. Colin},
  title   = {Some experiments on the recognition of speech, with one 
             and with two ears},
  journal = {Journal of the Acoustical Society of America},
  year    = {1953},
  volume  = {25},
  number  = {5},
  pages   = {975--979},
  doi     = {10.1121/1.1907229},
}

@article{geirnaert2021,
  author  = {Geirnaert, Simon and Francart, Tom and Bertrand, Alexander},
  title   = {Electroencephalography-based auditory attention decoding: 
             Toward neurosteered hearing devices},
  journal = {IEEE Signal Processing Magazine},
  year    = {2021},
  volume  = {38},
  number  = {4},
  pages   = {89--102},
  doi     = {10.1109/MSP.2021.3075932},
}

@article{biesmans2016,
  author  = {Biesmans, Wouter and Das, Neetha and Francart, Tom 
             and Bertrand, Alexander},
  title   = {Auditory-inspired speech envelope extraction methods for 
             improved {EEG}-based auditory attention detection in a 
             cocktail party scenario},
  journal = {IEEE Transactions on Neural Systems and Rehabilitation 
             Engineering},
  year    = {2017},
  volume  = {25},
  number  = {5},
  pages   = {402--412},
  doi     = {10.1109/TNSRE.2016.2571900},
}

@article{fuglsang2017,
  author  = {Fuglsang, S{\o}ren Asp and Dau, Torsten and 
             Hjortk{\ae}r, Jens},
  title   = {Noise-robust cortical tracking of attended speech in 
             real-world acoustic scenes},
  journal = {NeuroImage},
  year    = {2017},
  volume  = {156},
  pages   = {435--444},
  doi     = {10.1016/j.neuroimage.2017.04.026},
}

@misc{nju2023,
  author       = {Zhang, Yuanming and Yuan, Ziyan and Lu, Jing},
  title        = {{NJU} Auditory Attention Decoding Dataset},
  howpublished = {IEEE Dataport},
  year         = {2023},
  doi          = {10.21227/31nb-0j75}
}

@INPROCEEDINGS{esaa2022,
  author={Li, Peiwen and Su, Enze and Li, Jia and Cai, Siqi and Xie, Longhan and Li, Haizhou},
  booktitle={25th Conference of the Oriental COCOSDA International Committee for the Co-ordination and Standardisation of Speech Databases and Assessment Techniques (O-COCOSDA)}, 
  title={{ESAA}: An {EEG}-Speech Auditory Attention Detection Database}, 
  year={2022},
  volume={},
  number={},
  pages={1-6},
  doi={10.1109/O-COCOSDA202257103.2022.9997944}}

@article{mmaad2025,
  author  = {Fan, Cunhang and Zhang, Hongyu and Ni, Qinke and 
             Zhang, Jingjing and Tao, Jianhua and Zhou, Jian and 
             Yi, Jiangyan and Lv, Zhao and Wu, Xiaopei},
  title   = {Seeing helps hearing: {A} multi-modal dataset and a 
             {Mamba}-based dual branch parallel network for auditory 
             attention decoding},
  journal = {Information Fusion},
  year    = {2025},
  volume  = {118},
  pages   = {},
  doi     = {10.1016/j.inffus.2025.102946},
}

@article{rotaru2024,
  author  = {Rotaru, Iustina and Geirnaert, Simon and Heintz, Nathalie 
             and Van de Ryck, Iris and Bertrand, Alexander and 
             Francart, Tom},
  title   = {What are we really decoding? {Unveiling} biases in 
             {EEG}-based decoding of the spatial focus of auditory 
             attention},
  journal = {Journal of Neural Engineering},
  year    = {2024},
  volume  = {21},
  number  = {1},
  pages   = {},
  doi     = {10.1088/1741-2552/ad2214},
}

@article{golumbic2013,
  author  = {Zion Golumbic, Elana M. and Ding, Nai and Bickel, Stephan and Lakatos, Peter and Schevon, Catherine A. and McKhann, Guy M. and Goodman, Robert R. and Emerson, Ronald and Mehta, Ashesh D. and Simon, Jonathan Z. and Poeppel, David and Schroeder, Charles E.},
  title   = {Mechanisms underlying selective neuronal tracking of attended speech at a ``cocktail party''},
  journal = {Neuron},
  year    = {2013},
  volume  = {77},
  number  = {5},
  pages   = {980--991}
}

@article{hendrikse2019,
  author  = {Hendrikse, Maartje M. E. and Llorach, Gerard and Hohmann, Volker and Grimm, Giso},
  title   = {Movement and gaze behavior in virtual audiovisual listening environments resembling everyday life},
  journal = {Trends in Hearing},
  year    = {2019},
  volume  = {23},
  pages   = {}
}

@article{osullivan2015,
  author  = {O'Sullivan, James A. and Power, Alan J. and Mesgarani, Nima and Rajaram, Siddharth and Foxe, John J. and Shinn-Cunningham, Barbara G. and Slaney, Malcolm and Shamma, Shihab A. and Lalor, Edmund C.},
  title   = {Attentional selection in a cocktail party environment can be decoded from single-trial {EEG}},
  journal = {Cerebral Cortex},
  year    = {2015},
  volume  = {25},
  number  = {7},
  pages   = {1697--1706}
}

@article{ciccarelli2019,
  author  = {Ciccarelli, Gregory and Nolan, Michael and Perricone, Joseph and Calamia, Paul T. and Haro, Stephanie and O'Sullivan, James and Mesgarani, Nima and Quatieri, Thomas F. and Smalt, Christopher J.},
  title   = {Comparison of two-talker attention decoding from {EEG} with nonlinear neural networks and linear methods},
  journal = {Scientific Reports},
  year    = {2019},
  volume  = {9},
  number  = {1},
  pages   = {}
}

@inproceedings{lin2024,
  author    = {Zijie Lin and Tianyu He and Siqi Cai and Haizhou Li},
  title     = {{ASA}: An Auditory Spatial Attention Dataset with Multiple Speaking Locations},
  booktitle = {Interspeech 2024},
  pages     = {437--441},
  year      = {2024},
  doi       = {10.21437/Interspeech.2024-753}
}

@article{accou2023decoding,
  title={Decoding of the speech envelope from {EEG} using the {VLAAI} deep neural network},
  author={Accou, Bernd and Vanthornhout, Jonas and hamme, Hugo Van and Francart, Tom},
  journal={Scientific Reports},
  volume={13},
  number={1},
  pages={},
  year={2023},
  publisher={Nature Publishing Group UK London}
}

@inproceedings{accou2021modeling,
  title={Modeling the relationship between acoustic stimulus and {EEG} with a dilated convolutional neural network},
  author={Accou, Bernd and Monesi, Mohammad Jalilpour and Montoya, Jair and Francart, Tom and others},
  booktitle={European Signal Processing Conference (EUSIPCO)},
  pages={1175--1179},
  year={2021},
  organization={IEEE}
}

@article{accou2021predicting,
  title={Predicting speech intelligibility from {EEG} in a non-linear classification paradigm},
  author={Accou, Bernd and Jalilpour Monesi, Mohammad and Van Hamme, Hugo and Francart, Tom},
  journal={Journal of Neural Engineering},
  volume={18},
  number={6},
  pages={},
  year={2021},
  publisher={IOP Publishing}
}

@article{best2023effect,
  title={An effect of gaze direction in cocktail party listening},
  author={Best, Virginia and Boyd, Alex D and Sen, Kamal},
  journal={Trends in Hearing},
  volume={27},
  pages={},
  year={2023},
  publisher={SAGE Publications Sage CA: Los Angeles, CA}
}

@article{gehmacher2024eye,
  title={Eye movements track prioritized auditory features in selective attention to natural speech},
  author={Gehmacher, Quirin and Schubert, Juliane and Schmidt, Fabian and Hartmann, Thomas and Reisinger, Patrick and R{\"o}sch, Sebastian and Schwarz, Konrad and Popov, Tzvetan and Chait, Maria and Weisz, Nathan},
  journal={Nature Communications},
  volume={15},
  number={1},
  pages={},
  year={2024},
  publisher={Nature Publishing Group UK London}
}

@article{lertpoompunya2024head,
  title={Head-orienting behaviors during simultaneous speech detection and localization},
  author={Lertpoompunya, Angkana and Ozmeral, Erol J and Higgins, Nathan C and Eddins, David A},
  journal={Frontiers in Psychology},
  volume={15},
  pages={},
  year={2024},
  publisher={Frontiers Media SA}
}

@article{wallach1940role,
  title={The role of head movements and vestibular and visual cues in sound localization.},
  author={Wallach, Hans},
  journal={Journal of Experimental Psychology},
  volume={27},
  number={4},
  pages={339--368},
  year={1940},
  publisher={American Psychological Association}
}

@article{sumby1954visual,
  title={Visual contribution to speech intelligibility in noise},
  author={Sumby, William H and Pollack, Irwin},
  journal={Journal of the Acoustical Society of America},
  volume={26},
  number={2},
  pages={212--215},
  year={1954},
  publisher={Acoustical Society of America}
}

@article{ahmed2023integration,
  title={The integration of continuous audio and visual speech in a cocktail-party environment depends on attention},
  author={Ahmed, Farhin and Nidiffer, Aaron R and O'Sullivan, Aisling E and Zuk, Nathaniel J and Lalor, Edmund C},
  journal={NeuroImage},
  volume={274},
  pages={},
  year={2023},
  publisher={Elsevier}
}

@article{hurst2024gpt,
  title={Gpt-4o system card},
  author={Hurst, Aaron and Lerer, Adam and Goucher, Adam P and Perelman, Adam and Ramesh, Aditya and Clark, Aidan and Ostrow, AJ and Welihinda, Akila and Hayes, Alan and Radford, Alec and others},
  journal={arXiv preprint arXiv:2410.21276},
  year={2024}
}

@article{xu2026utilizing,
  title={Utilizing eyeblink information to improve {EEG}-based auditory attention decoding},
  author={Xu, Xiran and Xiao, Boda and Wang, Bo and Yan, Yujie and Wu, Xihong and Cheng, Heping and Chen, Jing},
  journal={Biomedical Signal Processing and Control},
  volume={123},
  pages={},
  year={2026},
  publisher={Elsevier}
}

@article{dimitrijevic2019neural,
  title={Neural indices of listening effort in noisy environments},
  author={Dimitrijevic, Andrew and Smith, Michael L and Kadis, Darren S and Moore, David R},
  journal={Scientific Reports},
  volume={9},
  number={1},
  pages={},
  year={2019},
  publisher={Nature Publishing Group UK London}
}

@article{seifi2020exploratory,
  title={An exploratory study of {EEG} alpha oscillation and pupil dilation in hearing-aid users during effortful listening to continuous speech},
  author={Seifi Ala, Tirdad and Graversen, Carina and Wendt, Dorothea and Alickovic, Emina and Whitmer, William M and Lunner, Thomas},
  journal={PLOS One},
  volume={15},
  number={7},
  pages={},
  year={2020},
  publisher={Public Library of Science San Francisco, CA USA}
}

@book{pearsons1977speech,
  title={Speech levels in various noise environments},
  author={Pearsons, Karl S and Bennett, Ricarda L and Fidell, Sanford A},
  year={1977},
  publisher={Office of Health and Ecological Effects, Office of Research and Development, U.S. Environmental Protection Agency.}
}

@article{smeds2015estimation,
  title={Estimation of signal-to-noise ratios in realistic sound scenarios},
  author={Smeds, Karolina and Wolters, Florian and Rung, Martin},
  journal={Journal of the American Academy of Audiology},
  volume={26},
  number={2},
  pages={183--196},
  year={2015},
  publisher={Thieme Medical Publishers}
}

@article{jaha2020visual,
  title={Visual Enhancement of Relevant Speech in a `{C}ocktail {P}arty'},
  author={Jaha, Niti and Shen, Stanley and Kerlin, Jess R and Shahin, Antoine J},
  journal={Multisensory Research},
  volume={33},
  number={3},
  pages={277--294},
  year={2020},
  publisher={Brill}
}

@article{wang2026open,
  title   = {An Open Non-Invasive {EEG} Dataset for Spontaneous Auditory Attention Switch Decoding},
  author  = {Wang, Xuefei and Ding, Yuting and Ban, Yueting and Wang, Lei and Chen, Fei},
  journal = {Scientific Data},
  volume  = {13},
  number  = {1},
  pages   = {},
  year    = {2026},
  doi     = {10.1038/s41597-026-07244-w},
  publisher = {Nature Publishing Group UK London}
}

@ARTICLE{pan10683957,
  author={Pan, Zexu and Borsdorf, Marvin and Cai, Siqi and Schultz, Tanja and Li, Haizhou},
  journal={IEEE/ACM Transactions on Audio, Speech, and Language Processing}, 
  title={{N}euro{H}eed: Neuro-Steered Speaker Extraction Using {EEG} Signals}, 
  year={2024},
  volume={32},
  number={},
  pages={4456-4470},
  doi={10.1109/TASLP.2024.3463498}}

@inproceedings{pan2024neuroheed+,
  title={{N}euro{H}eed+: Improving neuro-steered speaker extraction with joint auditory attention detection},
  author={Pan, Zexu and Wichern, Gordon and Germain, Fran{\c{c}}ois G and Khurana, Sameer and Le Roux, Jonathan},
  booktitle={IEEE International Conference on Acoustics, Speech and Signal Processing (ICASSP)},
  pages={11456--11460},
  year={2024},
  organization={IEEE}
}

@article{wang2020robust,
  title={Robust {EEG}-based decoding of auditory attention with high-rms-level speech segments in noisy conditions},
  author={Wang, Lei and Wu, Ed X and Chen, Fei},
  journal={Frontiers in Human Neuroscience},
  volume={14},
  pages={},
  year={2020},
  publisher={Frontiers Media SA}
}

@article{yang2026deep,
  title={Deep learning in auditory attention decoding: A systematic review},
  author={Yang, Jiashu and Huang, Mengjie and Yang, Rui},
  journal={Systems Science \& Control Engineering},
  volume={14},
  number={1},
  pages={},
  year={2026},
  publisher={Taylor \& Francis}
}

@inproceedings{farneback2003two,
  title={Two-frame motion estimation based on polynomial expansion},
  author={Farneb{\"a}ck, Gunnar},
  booktitle={Scandinavian Conference on Image Analysis},
  pages={363--370},
  year={2003},
  organization={Springer}
}

@article{crosse2016multivariate,
  title={The multivariate temporal response function (m{TRF}) toolbox: a {MATLAB} toolbox for relating neural signals to continuous stimuli},
  author={Crosse, Michael J and Di Liberto, Giovanni M and Bednar, Adam and Lalor, Edmund C},
  journal={Frontiers in Human Neuroscience},
  volume={10},
  pages={},
  year={2016},
  publisher={Frontiers Media SA}
}

@ARTICLE{heq1420370,
  author={de la Torre, A. and Peinado, A.M. and Segura, J.C. and Perez-Cordoba, J.L. and Benitez, M.C. and Rubio, A.J.},
  journal={IEEE Transactions on Speech and Audio Processing}, 
  title={Histogram equalization of speech representation for robust speech recognition}, 
  year={2005},
  volume={13},
  number={3},
  pages={355-366},
  doi={10.1109/TSA.2005.845805}}

@article{oganian2019speech,
  title={A speech envelope landmark for syllable encoding in human superior temporal gyrus},
  author={Oganian, Yulia and Chang, Edward F},
  journal={Science Advances},
  volume={5},
  number={11},
  pages={},
  year={2019},
  publisher={American Association for the Advancement of Science}
}

@misc{neurotoken2026,
  author       = {Alavi, Ali and Williamson, Donald},
  title        = {{NEUROTOKEN}: Joint Source and Directional {AAD} with Envelope Decoding via Conditional Flow Matching},
  howpublished = {},
  year         = {2026},
  note         = {Under review.}
}

 




\vfill

\end{document}